\documentclass[a4paper,12pt]{article}
\usepackage{amsmath}
\usepackage{amssymb}
\usepackage{amsfonts}
\usepackage{amssymb,amsmath}
\usepackage{cancel}
\usepackage{graphicx}
\usepackage{epsfig}
\usepackage{verbatim}
\usepackage{fleqn}
\usepackage[footnotesize]{caption}
\usepackage{mathrsfs}
\usepackage{color}
\usepackage{float}
\def\beq{\begin{equation}}
\def\eeq{\end{equation}}
\def\bea{\begin{eqnarray}}
\def\eea{\end{eqnarray}}

\def\<{\left\langle}
\def\>{\right\rangle}

\newcommand{\bc}{\begin{center}}
\newcommand{\ec}{\end{center}}
\newcommand{\bd}{\begin{displaymath}}
\newcommand{\ed}{\end{displaymath}}
\newcommand{\be}{\begin{equation}}
\newcommand{\ee}{\end{equation}}
\newcommand{\ba}{\begin{array}}
\newcommand{\ea}{\end{array}}
\newcommand{\bt}{\begin{tabular}}
\newcommand{\et}{\end{tabular}}

\newcommand{\BC}{\begin{center}}
\newcommand{\EC}{\end{center}}
\newcommand{\BE}{\begin{equation}}
\newcommand{\EE}{\end{equation}}
\newcommand{\BEA}{\begin{eqnarray}}
\newcommand{\BEAnn}{\begin{eqnarray*}}
\newcommand{\EEA}{\end{eqnarray}}
\newcommand{\EEAnn}{\end{eqnarray*}}

\newcommand{\VL}{\left( \begin{array}{c}}
\newcommand{\VR}{\end{array} \right)}
\newcommand{\ML}{\left( \begin{array}{cc}}
\newcommand{\MLd}{\left( \begin{array}{ccc}}
\newcommand{\MLv}{\left( \begin{array}{cccc}}
\newcommand{\MR}{\end{array} \right)}

\newcommand{\tsf}{\theta\kern-.20em_{\tilde{f}}}
\newcommand{\tsfp}{\theta\kern-.20em_{\tilde{f}\prime}}
\newcommand{\tsq}{\theta\kern-.15em_{\tilde{q}}}

\begin{document}

\bibliographystyle{OurBibTeX}

\begin{titlepage}

\vspace*{-25mm}
\begin{flushright}
UH511-1194-2012 \\ ADP-12-17/T784 \\ SHEP-12-07\\
\end{flushright}
%\vspace*{5mm}

\begin{center}
{ \sffamily \Large \bf %LHC constraints, reach and benchmarks of the
Constrained Exceptional Supersymmetric Standard Model with a Higgs Near 125 GeV
}
\\[8mm]
P.~Athron$^{a}$,
\footnote{E-mail: \texttt{peter.athron@adelaide.edu.au}}
S.F.~King$^{b}$,
\footnote{E-mail: \texttt{king@soton.ac.uk}}
D.J.~Miller$^{c}$,
\footnote{E-mail: \texttt{david.j.miller@glasgow.ac.uk}}
S.~Moretti$^{b}$
\footnote{E-mail: \texttt{stefano@phys.soton.ac.uk}}
and
R.~Nevzorov$^{d}$
\footnote{E-mail: \texttt{nevzorov@phys.hawaii.edu}.}
\footnote{On leave of absence from the Theory Department,
ITEP, Moscow, Russia.}
%\qquad\qquad\qquad\qquad\qquad\qquad\qquad
\\[3mm]
{\small\it
$^a$ ARC Centre of Excellence for Particle Physics at the Terascale, \\
School of Chemistry and Physics,
The University of Adelaide, \\
Adelaide, SA 5005, Australia. \\[2mm]
$^b$ School of Physics and Astronomy, University of Southampton,\\
Southampton, SO17 1BJ, U.K.\\[2mm]
$^c$ SUPA, School of Physics and Astronomy, University of Glasgow,\\
Glasgow, G12 8QQ, U.K.\\[2mm]
$^d$
Department of Physics and Astronomy, University of Hawaii, \\
Honolulu, HI 96822, U.S.A.
 }
\\[1mm]
\end{center}
\vspace*{0.5cm}

\begin{abstract}
\noindent
We study the parameter space of the constrained exceptional supersymmetric standard model (cE$_6$SSM)
consistent with a Higgs signal near 125 GeV and the LHC searches for squarks, gluinos and $Z^\prime$.
The cE$_6$SSM parameter space consistent with correct electroweak symmetry breaking,
is represented by scans in the $(m_0, M_{1/2})$ plane for fixed $Z'$ mass
and $\tan \beta$, with squark, gluino and Higgs masses plotted as contours in this plane.
We find that a 125 GeV Higgs mass only arises for a sufficiently large $Z'$ mass, mostly above current limits,
and for particular regions of squark and gluino masses corresponding to multi-TeV
squark masses, but with lighter gluinos typically within reach of
the LHC 8 TeV or forthcoming 14 TeV runs. Successful dark matter relic abundance may be achieved
over all the parameter space, assuming a bino-like LSP with a nearby heavier inert Higgsino doublet
and decoupled inert singlinos, resulting in conventional gluino decay signatures. A set of typical benchmark points
with a Higgs near 125 GeV is provided which exemplifies these features.
% and demonstrates the huge unexplored range of parameter space in this model with multi-TeV $Z'$ and squark %masses but with lighter gluinos, winos and binos, as well as possibly light coloured exotic $D$ fermions.
\end{abstract}

\end{titlepage}
\newpage
\section{Introduction}

 %\cite{ATLAS:2012ae}
  %\cite{Chatrchyan:2012tx}

  The ATLAS and CMS Collaborations have recently presented the first
indication for a Higgs boson with a mass about
$125$~GeV, consistent with the allowed window of Higgs masses
$125 \pm 3$ GeV \cite{ATLAS:2012ae,Chatrchyan:2012tx}.
In general, these results have generated much excitement in the
community, and already there are a number of papers discussing the
implications of such a Higgs boson
\cite{Higgs}. Many of these studies focus on the possibility that the
Higgs boson arises from a Supersymmetric Standard Model (SSM).
However there are several SSMs which are capable of giving rise to
a SM-like Higgs boson and it is interesting to survey some leading possibilities.
%A well known observation is that while the LHC can in principle
%exclude a Standard Model (SM) Higgs boson, it can only discover a
%SM-like Higgs boson, which could for example correspond to a SUSY
%Higgs boson near the decoupling region\cite{Haber}.

In the Minimal Supersymmetric Standard Model (MSSM) the lightest Higgs
boson is lighter than about 130-135 GeV, depending on top squark
parameters (see e.g.~\cite{Djouadi:2005gj} and references therein).
A 125 GeV SM-like Higgs boson is consistent with the MSSM in the
decoupling limit. In the limit of decoupling the light Higgs mass is given by
\begin{equation}
m_h^2 \, \approx  \, M_Z^2 \cos^2 2 \beta + \Delta m_h^2 \; ,
\label{eq:hmassMSSM}
\end{equation}
where $ \Delta m_h^2$ is dominated by loops of heavy top quarks and
top squarks and $\tan \beta$ is the ratio of the vacuum expectation
values (VEVs) of the two Higgs doublets introduced in the MSSM Higgs
sector. At large $\tan \beta$, we require $\Delta m_h \approx 85$~GeV
which means that a very substantial loop contribution, nearly as large
as the tree-level mass, is needed to raise the Higgs boson mass to 125
GeV.
%Due to the logarithmic dependence of the Higgs boson mass on the
%stop masses the latter depend exponentially on the Higgs boson mass.

%In this case the 125 GeV Higgs boson can be present in the particle spectrum
%even when the mixing in the stop sector is relatively small.
%The phenomenological implications of the NMSSM with extra exotic matter
%were considered in. \sn
%thereby constraining $\lambda \lesssim 0.7$ (everywhere in this paper
%$\lambda$ refers to the weak scale value of the coupling).

In the Next-to-Minimal Supersymmetric Standard Model (MSSM), the spectrum of the MSSM is extended by
one singlet superfield~\cite{genNMSSM1,genNMSSM2,Nevzorov:2004ge} (for reviews see
\cite{Ellwanger:2009dp}). In the NMSSM the
supersymmetric Higgs mass parameter $\mu$ is promoted to a
gauge-singlet superfield, $S$, with a coupling to the Higgs doublets,
$\lambda S H_u H_d$, that is perturbative up to unified scales.
In the pure NMSSM values of
$\lambda \sim 0.7$ do not spoil the validity of perturbation
theory up to the GUT scale only providing $\tan\beta\gtrsim 4$,
however the presence of additional extra matter allows smaller values of $\tan\beta$
to be achieved \cite{King:2012is}.
The maximum mass of the lightest Higgs boson in the NMSSM is
\begin{equation}
m_h^2 \, \approx  \,  M_Z^2 \cos^2 2 \beta + \lambda^2 v^2 \sin^2 2 \beta +  \Delta m_h^2 \;
\label{eq:hmassNMSSM}
\end{equation}
where here we use $v=174$~GeV.  For $\lambda v > M_Z$, the tree-level
contributions to $m_h$ are maximized for moderate values of $\tan \beta$
rather than by large values of $\tan \beta$ as in the MSSM. For example, taking
$\lambda =0.7$ and $\tan\beta=2$, these tree-level
contributions raise the Higgs boson mass to about 112 GeV,
and $\Delta m_h \gtrsim 55\,\mbox{GeV}$ is required. This is to be compared
to the MSSM requirement $\Delta m_h \gtrsim 85\,\mbox{GeV}$. The difference
between these two values (numerically about 30 GeV) is significant since
$\Delta m_h$ depends logarithmically on the stop
masses as well as receiving an important contribution from stop
mixing. This means for example, that, unlike the MSSM, in the case of
the NMSSM maximal stop mixing is not required to get the Higgs heavy
enough.

In the Exceptional Supersymmetric Standard Model (E$_6$SSM) \cite{King:2005jy,King:2005my},
the spectrum of the MSSM is extended to fill out three complete 27-dimensional representations
of the gauge group E$_6$ which is broken at the unification scale down to the SM gauge group
plus one additional gauged $U(1)_N$ symmetry at low energies under which right-handed neutrinos are neutral,
allowing them to get large masses.
Each $27$-plet contains one generation of ordinary matter; singlet fields, $S_i$; up and down type Higgs doublets, $H_{u,i}$ and $H_{d,i}$; charged $\pm 1/3$ coloured exotics $D_i$, $\bar{D}_i$.
The extra matter ensures anomaly cancellation, however
the model also contains two extra SU(2) doublets, $H'$ and $\bar{H}'$, which are required for gauge coupling unification \cite{King:2007uj}.
To evade rapid proton decay either a $Z_2^B$ or $Z_2^L$ symmetry is introduced and to evade large Flavour Changing Neutral Currents an approximate $Z_2^H$ symmetry is introduced which ensures that
only the third family of Higgs $H_{u,3}$ and $H_{d,3}$ couple to fermions and get vacuum expectation values (VEVs).
Similarly only the third family singlet $S_3$ gets a VEV, $\langle S_3 \rangle = s/\sqrt{2}$, which is responsible for the effective $\mu$ term and D-fermion mass.
The first and second families of Higgs and singlets which do not get VEVs are called ``inert''.
The maximum mass of the lightest SM-like Higgs boson in the E$_6$SSM is \cite{King:2005jy}
\begin{equation}
m_h^2 \, \approx  \,  M_Z^2 \cos^2 2 \beta + \lambda^2 v^2 \sin^2 2 \beta +
\frac{M_Z^2}{4} \left(1+ \frac{1}{4}\cos 2 \beta  \right)^2 + \Delta m_h^2 \;
\label{eq:hmassE6SSM}
\end{equation}
where the extra contribution relative to the NMSSM value
in Eq.~(\ref{eq:hmassNMSSM}) is due to the $U(1)_N$ D-term.
The Higgs mass can be larger due to two separate reasons, firstly the value of $\lambda$ may be larger
due to the extra matter, and secondly due to the $U(1)_N$ D-term contribution equal to
$\frac{1}{2}M_Z$  ($\frac{3}{8}M_Z$) GeV
for low (high) $\tan \beta$. For example for large $\tan \beta$, where the NMSSM term $\lambda^2 v^2 \sin^2 2 \beta$  is unimportant, the E$_6$SSM requires $\Delta m_h \approx 78$~GeV
as compared to $\Delta m_h \approx 85$~GeV in the MSSM.

In a previous paper we considered a constrained version of the E$_6$SSM with universal gaugino mass
$M_{1/2}$, soft scalar mass $m_0$, and soft trilinear mass $A$ at the unification scale $M_X$
\cite{Athron:2009bs}. Previous studies of the cE$_6$SSM have focussed on regions of cE$_6$SSM parameter space
which could have led to a discovery with the first LHC data \cite{Athron:2009ue} via the characteristic LHC signatures of the model \cite{Athron:2011wu}. These ``early'' benchmark points are by now excluded by LHC searches for SUSY and $Z'$ bosons. The main purpose of the present paper is to consider the cE$_6$SSM in the light of the Higgs signal near 125 GeV, taking into account the latest LHC constraints on squarks, gluinos and
$Z^\prime$ following the 7 TeV run. We find that there are huge unexplored regions of parameter space in the cE$_6$SSM which are consistent with a SM-like Higgs boson with a mass in the allowed window $125 \pm 3$ GeV.
The cE$_6$SSM parameter space consistent with correct electroweak symmetry breaking,
is represented here by scans in the $(m_0, M_{1/2})$ plane for fixed $Z'$ mass
and $\tan \beta$, with squark, gluino and Higgs masses plotted as contours in this plane.
If the Higgs mass is determined accurately then this will narrow down the
preferred regions of parameter space considerably.
For example, we find that a 125 GeV Higgs mass only arises for a sufficiently large $Z'$ mass, mostly above current limits,
and for particular regions of squark and gluino masses corresponding to multi-TeV
squark masses, but with lighter gluinos typically within reach of
forthcoming LHC 8 TeV or 14 TeV runs. Successful dark matter relic abundance may be achieved
over all the parameter space, assuming a bino-like LSP with a nearby heavier inert Higgsino doublet
and decoupled inert singlinos, resulting in conventional gluino decay signatures. A set of typical benchmark points
with a Higgs near 125 GeV is provided which exemplifies these features and demonstrates the huge unexplored range of parameter space in this model with multi-TeV $Z'$ and squark masses but with lighter gluinos, winos and binos, as well as possibly light coloured exotic $D$ fermions.

%%In addition we discuss how much more parameter space of the cE$_6$SSM the LHC will be able to explore. We then present and discuss the detailed phenomenology of new benchmark scenarios demonstarting the variety of signatures from the model which could be used to discover or further exclude the model.
The layout of the rest of the paper is as follows.
In section \ref{cE6SSM} we review the cE$_6$SSM.
In section \ref{constraints} we discuss existing LHC constraints arising from
Higgs searches, sparticle searches, exotica searches and $Z'$ searches.
In section \ref{DM} we show that successful dark matter relic abundance may be achieved
over all the parameter space, assuming a bino-like LSP with a nearby heavier inert Higgsino doublet
and decoupled inert singlinos, resulting in conventional gluino decay signatures.
In section \ref{Results} we provide detailed scans of the parameter space the cE$_6$SSM, presenting the results
in the $(m_0, M_{1/2})$ plane for fixed $Z'$ mass
and $\tan \beta$, with squark, gluino and Higgs masses plotted as contours in this plane. We also
present new heavy benchmarks for the model and discuss their phenomenology.
Section \ref{conclusions} concludes the paper.

%\newpage
\section{cE$_6$SSM \label{cE6SSM}}
The E$_6$SSM is a supersymmetric model based on the $SU(3)_C\times SU(2)_W\times U(1)_Y\times U(1)_N$
gauge group which is a subgroup of $E_6$. The extra $U(1)_N$ symmetry is the combination
$U(1)_{\chi}\cos\theta+U(1)_{\psi}\sin\theta$ with $\theta=\arctan\sqrt{15}$. In order to ensure
anomaly cancellation the particle content of the E$_6$SSM is extended to include three complete
fundamental $27$ representations of $E_6$. In addition the low energy particle spectrum of the E$_6$SSM
is supplemented by $SU(2)_W$ doublet $H'$ and anti-doublet $\overline{H}'$ states from
the extra $27'$ and $\overline{27'}$ to preserve gauge coupling unification. These components of
the $E_6$ fundamental representation originate from $\left(5^{*},\,2 \right)$ of $27'$
and $\left(5,\,-2 \right)$ of $\overline{27'}$ by construction. The analysis performed in \cite{King:2007uj}
shows that the unification of gauge couplings in the E$_6$SSM can be achieved for any phenomenologically
acceptable value of $\alpha_3(M_Z)$ consistent with the measured low energy central value, unlike in the
MSSM which, ignoring the effects of high energy threshold corrections, requires significantly higher
values of $\alpha_3(M_Z)$, well above the experimentally measured central value.
Because supermultiplets $H'$ and $\overline{H}'$ have opposite $U(1)_{Y}$ and $U(1)_{N}$ charges their
contributions to the anomalies are cancelled identically.

Thus, in addition to a $Z'$ associated with the
$U(1)_N$ symmetry, the E$_6$SSM involves extra matter beyond the MSSM with the quantum numbers of
three $5+5^{*}$ representations of $SU(5)$ plus three $SU(5)$ singlets with $U(1)_N$ charges.
The matter content of the E$_6$SSM with correctly normalized Abelian charges of all matter fields
is summarised in Table~\ref{charges}. The presence of a $Z'$ boson and exotic quarks predicted by
the E$_6$SSM provides spectacular new physics signals at the LHC which were discussed in
\cite{King:2005jy,King:2005my,Athron:2009bs,Athron:2009ue,Athron:2011wu,Accomando:2006ga}.

\begin{table}[ht]
\centering
\begin{tabular}{|c|c|c|c|c|c|c|c|c|c|c|c|c|c|}
\hline
 & $Q$ & $u^c$ & $d^c$ & $L$ & $e^c$ & $N^c$ & $S$ & $H_2$ & $H_1$ & $D$ &
 $\overline{D}$ & $H'$ & $\overline{H'}$ \\
 \hline
$\sqrt{\frac{5}{3}}Q^{Y}_i$
 & $\frac{1}{6}$ & $-\frac{2}{3}$ & $\frac{1}{3}$ & $-\frac{1}{2}$
& $1$ & $0$ & $0$ & $\frac{1}{2}$ & $-\frac{1}{2}$ & $-\frac{1}{3}$ &
 $\frac{1}{3}$ & $-\frac{1}{2}$ & $\frac{1}{2}$ \\
 \hline
$\sqrt{{40}}Q^{N}_i$
 & $1$ & $1$ & $2$ & $2$ & $1$ & $0$ & $5$ & $-2$ & $-3$ & $-2$ &
 $-3$ & $2$ & $-2$ \\
\hline
\end{tabular}
\caption{\it\small The $U(1)_Y$ and $U(1)_{N}$ charges of matter fields in the
    E$_6$SSM, where $Q^{N}_i$ and $Q^{Y}_i$ are here defined with the correct
$E_6$ normalisation factor required for the RG analysis.}
\label{charges}
\end{table}

Since right--handed neutrinos $N^c$ do not participate in gauge interactions
they are expected to gain masses at some intermediate scale after the breakdown
of $E_6$ \cite{King:2005jy,Howl:2008xz} while the remaining matter survives
down to the low energy scale near which the gauge group $U(1)_N$ is broken.
The heavy right--handed neutrinos shed light on the origin of the mass hierarchy in
the lepton sector allowing them to be used for the see--saw mechanism. At the same
time the heavy Majorana right-handed neutrinos may decay into final states with
lepton number $L=\pm 1$, thereby creating a lepton asymmetry in the early universe.
Since in the E$_6$SSM the Yukawa couplings of the new exotic particles are not
constrained by neutrino oscillation data, substantial values of the CP--asymmetries
can be induced even for a relatively small mass of the lightest right--handed neutrino
($M_1 \sim 10^6\,\mbox{GeV}$) so that successful thermal leptogenesis may be achieved
without encountering a gravitino problem \cite{King:2008qb}.

Although the presence of TeV scale exotic matter in E$_6$SSM gives rise to specatucular
collider signatures, it also leads to non--diagonal flavour transitions and rapid proton
decay. To suppress flavour changing processes as well as baryon and lepton number
violating operators one can postulate a $Z^{H}_2$ symmetry, under which all superfields
except one pair of $H_{d,i}$ and $H_{u,i}$ (say $H_d\equiv H_{d,3}$ and $H_u\equiv H_{u,3}$)
and one SM-type singlet superfield ($S\equiv S_3$) are odd \cite{King:2005jy,King:2005my}.
Here we also impose a discrete $Z^{S}_2$ symmetry, under which only first and second
generation singlet superfields are odd, i.e. $S_{1,2}\to -S_{1,2}$, whereas all other
supermultiplets are even \cite{Hall:2011zq}. In this case the fermionic components of $S_{1}$
and $S_{2}$ essentially decouple from the rest of the spectrum and the lightest neutralino
may be absolutely stable and can play the role of dark matter. The $Z^{H}_2$ and $Z^{S}_2$
symmetries reduce the structure of the Yukawa interactions to simplify the form of
the E$_6$SSM superpotential substantially. Integrating out heavy Majorana right--handed neutrinos
and keeping only Yukawa interactions whose couplings are allowed to be of order unity
leaves us with the following phenomenologically viable superpotential,
\begin{equation}
\ba{rcl}
W_{\rm E_6SSM}&\simeq &\lambda S (H_{d} H_{u})+\lambda_{\alpha}
S(H_{d,\alpha} H_{u,\alpha})+ \kappa_i S (D_i\overline{D}_i)\\[2mm]
&&+h_t(H_{u}Q)t^c+h_b(H_{d}Q)b^c+ h_{\tau}(H_{d}L)\tau^c+
\mu'(H^{'}\overline{H^{'}})+ h^{E}_{4j}(H_d H') e^c_j\,,
\ea
\label{cessm1}
\end{equation}
where $\alpha =1,2$ and $i=1,2,3$, and where the superfields $L=L_3$, $Q=Q_3$, $t^c=u^c_3$, $b^c=d^c_3$
and $\tau^c=e^c_3$ belong to the third generation and $\lambda_i$, $\kappa_i$ are dimensionless
Yukawa couplings with $\lambda \equiv \lambda_3$. In Eq.~(\ref{cessm1}) we choose the basis $H_{d,\alpha}$,
$H_{u,\alpha}$, $D_i$ and $\overline{D}_i$ so that the Yukawa couplings of the singlet field $S$
have flavour diagonal structure. Hereafter, we assume that the couplings $h^{E}_{4j}$ are rather small
and can be neglected.

From Eq.~(\ref{cessm1}) it follows that the $SU(2)_W$ doublets $H_u$ and $H_d$, that are even
under the $Z^{H}_2$ symmetry, play the role of Higgs fields generating the masses of quarks and leptons
through electroweak (EW) symmetry breaking (EWSB) while the other generations of these Higgs like
fields remain inert. The singlet field $S$ must also acquire a large VEV in order to induce sufficiently
large masses for the exotic charged fermions and $Z'$ boson. The couplings $\lambda_i$ and $\kappa_i$
should be large enough to ensure that the exotic fermions are sufficiently heavy to avoid conflict with
direct particle searches at present and former accelerators. If $\lambda_i$ or $\kappa_i$ are reasonably
large they affect the evolution of the soft scalar mass $m_S^2$ of the singlet field $S$ rather strongly
resulting in negative values of $m_S^2$ at low energies that triggers the breakdown of the $U(1)_{N}$ symmetry.

Initially the sector of EWSB in the E$_6$SSM involves ten degrees of freedom. However four of them are massless
Goldstone modes which are swallowed by the $W^{\pm}$, $Z$ and $Z'$ gauge bosons that gain non-zero masses.
If CP--invariance is preserved the other degrees of freedom form two charged, one CP--odd and three CP-even
Higgs states. When the SUSY breaking scale is considerably larger than the EW scale, the mass matrix of the
CP-even Higgs sector has a hierarchical structure and can be diagonalised using perturbation theory
\cite{Nevzorov:2004ge,Nevzorov:2001um}. In this case the mass of one CP--even Higgs particle is always
close to the $Z'$ boson mass $M_{Z'}$. The masses of another CP--even, the CP--odd and the charged Higgs
states are almost degenerate. When $\lambda\gtrsim g'_1$, the qualitative pattern of the Higgs spectrum is
rather similar to the one which arises in the PQ symmetric NMSSM \cite{Nevzorov:2004ge,Miller:2005qua}.
In the considered limit the heaviest CP--even, CP--odd and charged states are almost degenerate and lie beyond
the $\mbox{TeV}$ range \cite{King:2005jy}. Finally, like in the MSSM and NMSSM, one of the CP--even Higgs bosons
is always light irrespective of the SUSY breaking scale. However, in contrast with the MSSM, the lightest Higgs
boson in the E$_6$SSM can be heavier than $110-120\,\mbox{GeV}$ even at tree level. In the two--loop approximation
the lightest Higgs boson mass does not exceed $150-155\,\mbox{GeV}$ \cite{King:2005jy}. In our analysis here
we explore the part of the parameter space of the constrained E$_6$SSM which is associated with the SM--like
Higgs boson mass around $125\,\mbox{GeV}$.

Although $Z^{H}_2$ eliminates problems related with baryon number violation and non-diagonal flavour transitions
it also forbids all interactions that allow the lightest exotic quarks to decay. Since models with stable
charged exotic particles are ruled out by experiment \cite{42} the $Z^{H}_2$ symmetry can only be
approximate. The appropriate suppression of the proton decay rate can be achieved if one imposes either a $Z_2^L$
or a $Z_2^B$ discrete symmetry \cite{King:2005jy}. If the Lagrangian is invariant with respect to a $Z_2^L$ symmetry,
under which all superfields except lepton ones are even (Model I), then the terms in the superpotential which
permit the lightest exotic quarks to decay and are allowed by the gauge symmetry can be written as follows
\begin{equation}
W_1=g^Q_{ijk} D_{i} (Q_j Q_k)+
g^{q}_{ijk}\overline{D}_i d^c_j u^c_k\,.
\label{cessm2}
\end{equation}
In this case the baryon number conservation requires exotic quarks to be diquarks. The invariance of the Lagrangian
with respect to $Z_2^B$ symmetry (Model II), under which supermultiples $H^d_{i}$, $H^u_{i}$, $S_i$, $Q_i$, $u^c_i$, $d^c_i$
are even while the exotic quark ($D_i$ and $\overline{D_i}$) as well as lepton superfields ($L_i$, $e^c_i$, $N^c_i$)
are odd, implies that the following couplings are allowed:
\begin{equation}
W_2=g^E_{ijk} e^c_i D_j u^c_k+
g^D_{ijk} (Q_i L_j) \overline{D}_k\,.
\label{cessm3}
\end{equation}
As a consequence, in Model II, $\overline{D}_i$ and $D_i$ manifest themselves in the Yukawa interactions
as leptoquarks. With both of these symmetries the MSSM particle content behaves like it does under $R$--parity, with the
subset of particles present in the standard model and Higgs (and also inert Higgs) bosons being even under this generalised
$R$--parity, while their supersymmetric partners are odd and therefore, as usual, must be pair produced, and upon
decaying will always give rise to a stable lightest supersymmetric particle (LSP). However the exotic $D$-fermions are
odd and so must be pair produced and will decay into an LSP, while their scalar superpartners are even and in some
cases can be singly produced.

In both models discussed above the $Z^{H}_2$ symmetry violating couplings are not forbidden. Nevertheless because the
$Z^{H}_2$ symmetry violating operators lead to non--diagonal flavour interactions, the corresponding Yukawa couplings are
expected to be small, and must preserve either the $Z_2^B$ or $Z_2^L$ symmetry to ensure proton stability. In particular,
to suppress flavour changing processes the Yukawa couplings of the inert Higgs states to the quarks and leptons of the
first two generations should be smaller than $10^{-3}-10^{-4}$. In our analysis small $Z^{H}_2$ symmetry violating
couplings can be ignored in the first approximation.

Assuming that $h^{E}_{4j}\to 0$ the superpotential of the E$_6$SSM which is invariant with respect to both
$Z_2^H$ and $Z_2^S$ symmetries involves only six extra Yukawa couplings ($\lambda_i$ and $\kappa_i$) as compared with
the MSSM with $\mu=0$. The soft breakdown of SUSY gives rise to many new parameters. For instance, it induces additional
trilinear scalar couplings associated with the Yukawa interactions as well as a set of soft scalar masses.
The number of fundamental parameters reduces drastically within a constrained version of the model (cE$_6$SSM)
\cite{Athron:2009bs,Athron:2009ue}, defined at the GUT scale $M_X$, where all gauge couplings coincide, i.e.
$g_1(M_X) \simeq g_2(M_X)\simeq g_3(M_X) \simeq g'_1(M_X)$. Constrained SUSY models imply that all soft scalar masses
are set to be equal to $m_0$ at some high energy scale $M_X$, all gaugino masses $M_i(M_X)$ are equal to $M_{1/2}$ and
trilinear scalar couplings are such that $A_i(M_X)=A_0$. Thus the cE$_6$SSM is characterised by the following set of
Yukawa couplings and universal soft SUSY breaking terms,
\begin{equation}
\lambda_i(M_X),\quad \kappa_i(M_X),\quad h_t(M_X),\quad h_b(M_X), \quad h_{\tau}(M_X),
\quad m_0, \quad M_{1/2},\quad A_0,
\label{cessm4}
\end{equation}
where $h_t(M_X)$, $h_b(M_X)$ and $h_{\tau}(M_X)$ are the usual $t$--quark, $b$--quark and $\tau$--lepton Yukawa couplings,
and $\lambda_i(M_X)$, $\kappa_i(M_X)$ are the extra Yukawa couplings defined in Eq.~(\ref{cessm4}). Near the GUT scale $M_X$
the part of the scalar potential $V_{soft}$ that contains a set of the soft SUSY breaking terms takes the form
\begin{equation}
V_{soft}= m_0^227_i27_i^*+A_0Y_{ijk}27_i27_j27_k +h.c.,
\label{cessm5}
\end{equation}
where $Y_{ijk}$ are generic Yukawa couplings from the trilinear terms in Eq.~(\ref{cessm1}) and the $27_i$ represent
generic fields from Table~\ref{charges} and in particular those which appear in Eq.~(\ref{cessm1}). Since $Z^{H}_2$ and $Z^{S}_2$
symmetries forbid many terms in the superpotential of the E$_6$SSM it also forbids similar soft SUSY breaking terms in
$V_{soft}$. In order to guarantee correct EWSB, $m_0^2$ has to be positive.  To simplify our analysis we also assume that
$A_0$ is real and $M_{1/2}$ is positive --- this then naturally leads to real VEVs of the Higgs fields.

The set of cE$_6$SSM parameters in Eq.~(\ref{cessm4}) should in principle be supplemented by $\mu'$ and the associated
bilinear scalar coupling $B'$. The term $\mu'(H^{'}\overline{H^{'}})$ is the only bilinear term in the superpotential
Eq.~(\ref{cessm1}). It is solely responsible for the masses of the charged and neutral components of $H^{'}$ and $\overline{H^{'}}$.
The corresponding mass term is not suppressed by $E_6$ symmetry and is not involved in the process of EWSB. Therefore the
parameter $\mu'$ remains arbitrary. Recent analysis revealed that the gauge
coupling unification in the E$_6$SSM is consistent with $\mu'$ around $100\,\mbox{TeV}$ \cite{King:2007uj}.
Therefore we assume that the parameter $\mu'$ can be as large as $10\,\mbox{TeV}$ so that the scalar and fermion
components of the superfields $H'$ and $\overline{H'}$ are very heavy. As a result they decouple from the rest
of the particle spectrum and the parameters $B'$ and $\mu'$ are irrelevant for our analysis. This also justifies why the
Yukawa couplings $h^{E}_{4j}$ can be neglected in the first approximation if they are not too large.

To calculate the particle spectrum within the cE$_6$SSM one must find sets of parameters which are consistent with both
the high scale universality constraints and the low scale EWSB constraints. To evolve between these two scales we use
two--loop renormalisation group equations (RGEs) for the gauge and Yukawa couplings together with two--loop RGEs for
$M_a(Q)$ and $A_i(Q)$ as well as one--loop RGEs for $m_i^2(Q)$. The low energy values of the soft SUSY breaking
terms can be determined semi-analytically as functions of $A_0$, $M_{1/2}$ and $m_0$. The corresponding semi-analytic
expressions can be written as
\begin{equation}
\begin{array}{rcl}
m_i^2(Q) &=& a_i(Q)  M_{1/2}^2 + b_i(Q) A_0^2 + c_i(Q) A_0 M_{1/2} + d_i(Q) m_0^2,\\
A_i(Q) &=& e_i(Q) A_0 + f_i(Q) M_{1/2},\qquad\qquad M_i(Q) = p_i(Q) A_0 + q_i(Q) M_{1/2},
\end{array}
\label{cessm6}
\end{equation}
where $Q$ is the renormalisation scale. The analytic expressions for the coefficients
$a_i(Q)$, $b_i(Q)$, $c_i(Q)$, $d_i(Q)$, $e_i(Q)$, $f_i(Q)$, $p_i(Q)$, $q_i(Q)$
are unknown, since an exact analytic solution of the E$_6$SSM RGEs is not available.
Nevertheless these coefficients may be calculated numerically at the low energy scale.
We use semi-analytic expressions Eq.~(\ref{cessm6}) in our analysis to determine the sets of
$m_0$, $M_{1/2}$ and $A_0$ which are consistent with EWSB. This allows one to replace $m_0$, $M_{1/2}$
and $A_0$ by $v$, $\tan \beta$ and $s$ through the EWSB conditions, in a similar manner to the
way $|\mu|$ and $B$ are traded for $\tan\beta$ and $v$ in the MSSM. This means that the particle
spectrum and other phenomenological aspects of the model are defined by only eight free parameters,
which in previous analyses have been taken to be $\{ \lambda_i$, $\kappa_i$, $s$, $\tan \beta$\}
\footnote{This should be compared to the cMSSM with $\{m_0, M_{1/2}, A, \tan \beta,
{\rm sign}(\mu)$\}.} and can be reduced further by considering scenarios with some Yukawa coupling
universality or other well motivated relations between the Yukawa couplings at the GUT scale.

Although correct EWSB is not guaranteed in the cE$_6$SSM, remarkably, there are always solutions
with real $A_0$, $M_{1/2}$ and $m_0$ for sufficiently large values of $\kappa_i$, which drive
$m_S^2$ negative. This is easy to understand since the $\kappa_i$ couple the singlet to a large
multiplicity of coloured fields, thereby efficiently driving its squared mass negative to trigger
the breakdown of the gauge symmetry.

To calculate the particle spectrum within the cE$_6$SSM a private spectrum generator has been written,
based on some routines and the class structure of SOFTSUSY 2.0.5 \cite{Allanach:2001kg}. The details
of the procedure we followed, including the RGEs for the E$_6$SSM and the experimental and theoretical
constraints can be found in \cite{Athron:2009bs, Athron:2009ue}. To avoid any conflict with
collider experiments as well as with recent cosmological observations we impose
the set of constraints specified in the next section. These bounds restrict the allowed range of the
parameter space in the cE$_6$SSM.

%\newpage
\section{LHC constraints \label{constraints}}
\subsection{Higgs searches}
At present, the situation in ATLAS on Higgs mass limits within the SM hypothesis is well summarised in \cite{ATLAS:2012ae}.
Herein, a preliminary combination of SM Higgs boson searches was performed in a dataset corresponding to an integrated
luminosity of  4.6 to 4.9 fb$^{-1}$ taken at 7 TeV. A SM Higgs boson is excluded at the 95\% CL in the mass
ranges 110.0--117.5~GeV, 118.5--122.5 GeV and 129--539 GeV, while the range 120--555 GeV is
expected to be excluded in the absence of a signal. The mass regions between 130 and 486 GeV are excluded at the 99\% CL.
An excess of events is observed around 126 GeV with a local significance of $2.5\sigma$, where the expected significance in
the presence of a  SM Higgs boson for that mass hypothesis is $2.9\sigma$.

Combined results were reported by CMS in \cite{Chatrchyan:2012tx}, based on searches for the SM Higgs boson at 7 TeV in the usual
five decay modes: $\gamma\gamma$, $b\bar b$, $\tau^+\tau^-$, $WW$ and $ZZ$. The explored Higgs boson mass range is
110--600 GeV. The analysed data correspond
to an integrated luminosity of 4.6--4.8 fb$^{-1}$. The expected excluded mass range in the absence of the
SM Higgs boson is 118--543 GeV at 95\% CL. The observed results exclude the SM Higgs boson in the mass range
127--600 GeV at 95\% CL and in the mass range 129--525 GeV at 99\% CL. An excess of events above the expected SM background
is observed at the low end of the explored mass range making the observed limits weaker than expected in the absence of a
signal. The largest excess, with a local significance of  $3.1\sigma$, is observed for a SM Higgs boson mass hypothesis of
124 GeV.

All our benchmarks presume the lightest Higgs boson mass in the tentative signal range of 124--126 GeV. Further,
by making use of a modification of the programs described in Refs.~\cite{Moretti:1994ds,Kunszt:1996yp},
we have checked that the cross section times BR rates for the process $pp\to h_1\to X$, where $X$
represents the aforementioned channels in which ATLAS and CMS have shown sensitivity to a Higgs boson with such a mass,
as obtained in the cE$_6$SSM, differ by no more than $7-8\%$ from those of the SM.
In particular, notice that we have allowed in the case of the cE$_6$SSM also for all possible non-SM states belonging to its
spectrum that could enter the $gg\to h_1$ loop diagram at production level and the $h_1\to gg,\gamma\gamma,Z\gamma$ loop
diagrams at decay level. Masses of the SM states were the same in both calculations, while the relevant gauge couplings
were different, as extracted from the corresponding RGEs of the two models. 

\subsection{Sparticle searches \label{sparticle_searches}}

Recent searches for supersymmetry by both ATLAS and CMS have considerably reduced the available parameter space for low energy supersymmetric models. Of particular interest are searches for squarks of the first two generations and gluinos, which have been probed by ATLAS using final states with jets and missing transverse momentum and possibly an isolated lepton \cite{ATLAS_susy1,ATLAS_susy2,ATLAS_susy3}, all performed with $4.7\,{\rm fb}^{-1}$ of data. Similarly, the CMS collaboration has provided interesting exclusions by forcing events with missing transverse energy into a dijet topology \cite{CMS_razor} using $4.4\,{\rm fb}^{-1}$ of data, or alternatively by using the $M_{T2}$ variable \cite{CMS_MT2} on $4.73\,{\rm fb}^{-1}$ of data.

These exclusions are, of course, sensitive to the details of the supersymmetric model.  Both ATLAS and CMS chose to interpret their searches as exclusions in the $m_0-M_{1/2}$ plane of the cMSSM fixing values for the other cMSSM parameters $\tan \beta$, $A_0$ and the sign of $\mu$. For our purposes, these exclusions must be reinterpreted for the cE$_6$SSM. This presents two difficulties. First of all, the $m_0$ and $M_{1/2}$ values of the cMSSM bear little relation to their counterparts in the cE$_6$SSM; the RGE running from the unification scale results in a completely different low energy spectrum, so a particular choice of $m_0$ and $M_{1/2}$ will yield different squark and gluino masses in each model. This is further exacerbated by the arbitrary choice of parameters $\tan \beta$, $A_0$ and the sign of $\mu$, which will not, in general, correspond to the parameter choices for the cE$_6$SSM. Fortunately, both CMS and ATLAS have also superimposed contours of squark and gluino masses on their exclusion plots. These contours tell us that for reasonably heavy squarks, above about $1.5\,$TeV, we must ensure that our gluinos are heavier than about $850\,$GeV or so.

The ``about'' and ``or so'' of the last sentence is a product of our second difficulty: the squarks and gluinos must necessarily decay to lighter supersymmetric states in the spectrum, including the lightest supersymmetric particle (LSP). As already pointed out, the $m_0$ and $M_{1/2}$ used to determine the (cMSSM) experimental exclusions may predict a rather different spectrum for the cE$_6$SSM scenario with analogous squark and gluino masses. Therefore the cE$_6$SSM squarks and gluinos may have decay widths and branching ratios considerably different from those used for the experimental exclusion. It has been pointed out in \cite{LeCompte:2011cn} that supersymmetry with a {\it compressed} specturm, that is with smaller mass intervals between the particle states, may avoid experimental exclusions since the decays may contain less missing momentum.

Without a dedicated experimental analysis of the E$_6$SSM we are unable to determine how the experimental limits on squarks and gluinos will be changed by these effects. Consequently we will adopt a conservative approach and insist that our benchmark scenarios are considerably beyond these limits; namely that our first and second generation squarks are significantly heavier than $1.4\,$TeV and our gluinos are significantly heavier than $850\,$GeV.

Searches for the third generation squarks by ATLAS are less well developed and as, as yet, only available for the $2.1\,{\rm fb}^{-1}$ dataset. A search for bottom squarks in the MSSM, assuming a $100\%$ branching ratio for the decay $\tilde b_1 \to b \tilde \chi_1^0$,  excludes bottom squark masses up to $390\,$GeV for neutralino masses below $60\,$GeV \cite{Aad:2011cw}. Top squark constraints are also rather weak, with an exclusion of top squark masses below $310\,$GeV as long as the neutralino is in the mass window $115-230\,$GeV \cite{ATLAS_stop}. This study assumed a GMSB model.

CMS has also produced an exclusion in the plane of the gluino and LSP masses for a simplified model \cite{CMS_MT2,Alves:2011wf} using the process $pp \to \tilde g \tilde g$ with $\tilde g \to bb \chi_1^0$ and $4.73\,{\rm fb}^{-1}$ (also see \cite{CMS_simp}). For gluino masses below about $1\,$TeV, this analysis excludes a lightest neutralino  lighter than about $440\,$GeV, with this limit reducing quickly for higher gluino masses, disappearing entirely by  $1.06\,$TeV. However, this simplified model requires a rather specific spectrum, and it is not clear how applicable this is to our cE$_6$SSM case.

One final analysis of note is an ATLAS search for direct neutralino and chargino production in a simplified model where the lightest chargino and next-to-lightest neutralino are degenerate \cite{ATLAS_chargino}. Using $2.06\,{\rm fb}^{-1}$ of data, this study concluded that these degenerate $\tilde \chi_1^{\pm} / \tilde \chi_2^0$ are excluded up to $300\,$GeV for a lightest neutralino lighter than $250\,$GeV.

\subsection{Exotica searches}
\label{exotic-searches}
The production of a TeV scale exotic colored states should also lead to spectacular LHC signals.
Several experiments at LEP, HERA, Tevatron and LHC were searching for the colored objects
that decay into either a pair of quarks or quark and lepton. But most searches imply that
exotic color states, i.e leptoquarks or diquarks, have integer--spin. So they are either
scalars or vectors. Because of this new colored objects can be coupled directly to either
a pair of quarks or to quark and lepton. Moreover it is usually assumed that leptoquarks and
diquarks have appreciable couplings to the quarks and leptons of the first generation.
The most stringent constraints on the masses of leptoquarks come from the non-observation
of these exotic color states at the ATLAS experiment. Recently ATLAS collaboration ruled out
first and second generation scalar leptoquarks (i.e. leptoquarks that couple to the first and
second generation fermions respectively) with masses below $320-420\,\mbox{GeV}$
\cite{Aad:2011uv}. The experimental lower bounds on the masses of diquarks (dijet resonances)
tend to be considerably higher (see, for example, \cite{90}).

However the LHC lower bounds on the masses of exotic colored states mentioned above are 
not directly applicable in the case of the E$_6$SSM (also see \cite{Nevzorov:2012hs}). 
Indeed, our analysis of the particle spectrum within cE$_6$SSM indicates that $\tilde{D}$-scalars 
tend to be rather heavy. On the other hand exotic $D$--fermions can have masses below the 
TeV scale. Assuming that they couple most strongly to the third family (s)quarks and (s)leptons, 
the lightest exotic $D$--fermions decay into $\tilde{t}b$, $t\tilde{b}$, $\bar{\tilde{t}}\bar{b}$, 
$\bar{t}\bar{\tilde{b}}$ (if they are diquarks) or $\tilde{t}\tau$, $t\tilde{\tau}$, 
$\tilde{b} \nu_{\tau}$, $b\tilde{\nu_{\tau}}$ (if they are leptoquarks) resulting in the missing 
energy and transverse momentum in the final state. This is because these states are odd under 
the analogue of $R$--parity in the E$_6$SSM. Due to the presence of missing energy in the 
final states of the decays of $D$--fermions a special dedicated study is required to determine 
the experimental limits on their masses and couplings.

\subsection{$Z'$ searches}
Recent 95\% CL mass limits on $Z'$ bosons of $E_6$ origin (in di-lepton searches) from ATLAS based on 5 fb$^{-1}$
of data collected at 7 TeV were reported in \cite{ATLAS-Zp}, where
a  lower limit of 2.21 TeV on the mass of the Sequential Standard Model (SSM) $Z'$ boson is set, re-scalable to $1.78$ TeV in the case of a $Z'_N$ boson.
Limits were also reported by CMS in \cite{CMS-Zp} for 1.1 fb$^{-1}$ of luminosity but have now been superseded by one at full luminosity \cite{:2012it} which has just been announced as this paper is finalised, setting a lower limit of 2.08 TeV on the mass of the $Z'_N$ boson.

However the limits quoted by ATLAS and CMS are for the specified  $Z^\prime$ couplings with the assumption that the decay into SM particles provides the only
kinematically
allowed decay channels.   In \cite{Athron:2011wu} the impact of exotics decay width was studied and a considerable impact was found for two test case benchmarks.  Based on the work in \cite{Accomando:2010fz} we then used the branching ratios for those benchmarks (which reduced the branching into leptons, compared to ignoring exotic decays, by about a factor two) to rescale the cross section prediction and obtained an estimate of how the limit can changes when light exotics, if present, are taken into.  For example if we assume a similar impact from exotics (i.e. a dilution of the leptonic branching ratio by a factor two) then the limit on the $Z^\prime_N$ mass could be reduced from $2.02$ TeV to somewhere around $1.8$ TeV, as can be seen from examining figure 6 of  \cite{:2012it}.

However one should note that not only is this a fairly simple estimate, and is also dependent on the details of the spectrum and the masses of the various exotic states.  However the first two generations of singlinos had a significant contribution to the width and light singlinos are always present, therefore the limit will always be significantly smaller than that quoted by assuming no available exotic decay channels.  

 The benchmarks and plots presented here all have a $Z'$ mass above $1.8$ TeV and in all but one plot and benchmark (where $M_{Z^\prime} = 1.889$ TeV) also above the quoted limit assuming no exotics and therefore clearly safe in this respect.

\section{Dark Matter constraints\label{DM}}

We now consider the question of cosmological cold dark matter (CDM) relic abundance due to the neutralino LSP.
In the considered benchmark points we have a predominantly
bino-like lightest neutralino with a mass $|m_{\chi_1^0}|$.
One might be concerned that such a bino
might give too large a contribution to $\Omega_{CDM}$. Indeed a recent
calculation of $\Omega_{CDM}$ in the USSM \cite{Kalinowski:2008iq},
which includes the effect of the MSSM states plus the extra $Z'$ and
the active singlet $S$, together with their superpartners, indicates
that for the benchmarks considered here that $\Omega_{CDM}$ would be
too large.  However the USSM does not include the effect of the extra
inert Higgs and Higgsinos that are present in the E$_6$SSM, and so we need to discuss their effect
on the relic density.

In our analysis we have considered the inert Higgsino masses given by
$\mu_{\tilde{H}} (\alpha ) = \lambda_{\alpha} s/\sqrt{2}$. We have not
considered the mass of the inert singlinos. In general, these are
generated by mixing with the Higgs and inert Higgsinos, and are thus
of order $fv^2/s$, where $f$ are additional Yukawa couplings that we
have not specified in our analysis. Since $s\gg v$ it is quite likely
that the LSP neutralino in the cE$_6$SSM will be an inert singlino
with a mass lighter than 60 GeV. This would imply that the state
$\chi_1^0$ considered here is not cosmologically stable but would
decay into lighter states consisting of admixtures of inert singlinos
and inert higgsinos. Such states can annihilate via an s-channel
Z-boson, due to their doublet component, yielding an acceptable CDM
relic abundance, as has been recently been demonstrated in the
E$_6$SSM \cite{Hall:2009aj}.  However, in such a scenario, the
lightest inert neutralino states would have sizable couplings to the
Higgs boson, leading to significant modifications of Higgs
phenomenology, and a contribution to the direct detection
cross-section to due Higgs exchange which conflicts with the XENON-100
limit in the region of parameter space where the correct relic
abundance is achieved \cite{Hall:2010ix}. There are several ways out
of this problem, as discussed in \cite{Belyaev:2012si}. For example,
the light inert states may have masses around the keV and GeV energy
scale, evading the XENON-100 search limits, and leading to Warm Dark
Matter \cite{King:2012wg}.  Many of these possibilities lead to novel
gluino decays into the extra lighter neutralinos and charginos, which
can modify significantly the gluino search limits and strategies as
discussed in \cite{Belyaev:2012si}.

In this paper we shall consider the approach to CDM in the E$_6$SSM
proposed in \cite{Hall:2011zq}.  The idea is to set the inert singlino
$\tilde{S}_{1,2}$ couplings $f$ to zero so that they are exactly
massless, with their couplings forbidden by a discrete symmetry. The
massless $\tilde{S}_{1,2}$ singlinos will contribute to the effective
number of neutrino species in the Early Universe giving
$N_{eff}\approx 3.2$, but will otherwise be unobservable, except in
$Z'$ decays. In particular they play no part in dark matter. Thus, the
bino-like state with a mass $|m_{\chi_1^0}|$ will be cosmologically
stable and will be a CDM candidate. In order to achieve the observed
WMAP relic abundance, we shall tune the mass of the lightest inert
Higgsino mass $\mu_{\tilde{H}} (1) = \lambda_{1} s/\sqrt{2}$ to be
just above the bino mass. The idea is that the bino up-scatters into
the nearby inert Higgsinos which subsequently efficiently annihilate
via a $Z$ boson into SM particles \cite{Hall:2011zq}. Note that the
inert Higgsinos have full electroweak strength couplings to the $Z$
boson.  In practice the correct relic abundance can be achieved by
tuning the inert Higgsinos to be about 10 GeV heavier than the bino
\cite{Hall:2011zq},
\begin{equation}
\mu_{\tilde H} (1) \approx |m_{\chi_1^0}|+10\ GeV .
\label{DM-cond}
\end{equation}

In the parameter space scans discussed later we do not directly impose
this condition, since we do not know the bino mass at the outset.
However it is always possible to satisfy Eq.~(\ref{DM-cond}) by tuning
$\lambda_{1}(M_X)$, the Yukawa coupling which fixes the lightest inert
Higgsino mass.  Additionally $\lambda_{2}(M_X)$ (Yukawa coupling for
the heaviest inert Higgsino) can be tuned to compensate the
impact on the RGEs such that the rest of the mass spectrum is
unchanged.  Therefore Eq.~(\ref{DM-cond}) can be satisfied for every
point on all of the plots shown.  All benchmarks presented will also be
required to satisfy this condition.

We emphasise that, in this scenario, the gluino decays will be just those of the MSSM with
a bino-like LSP, so standard MSSM gluino searches and limits will also apply to the E$_6$SSM.

\section{Results \label{Results}}

\subsection{Exploration of the parameter space}

The LHC limits coming from searches for a $Z^\prime$, squarks and
gluinos and the Higgs restrict the parameter space of the model in
complementary ways. Additionally the tentative signal Higgs signal
between 124 -- 126 GeV, if confirmed and the mass precisely measured,
would also substantially improve our knowledge of the cE$_6$SSM
parameters. One should also note that as described in
\cite{King:2005jy, King:2005my} studies based on the (unconstrained)
E$_6$SSM showed that the light Higgs can be substantially heavier in
the E$_6$SSM than in the MSSM so there is significant reason for
optimism that this signal could be comfortably accommodated in this
model.

To explore this further we carried out a number of scans over the parameter space. In each scan we fixed $s$ (and therefore the $Z^\prime$ mass) so that the $Z^\prime$ mass is above the experimental limit.  For scenarios where the $Z^\prime$ is below its mass limits, most of the parameters space would also have gluinos below the limit suggested by the LHC searches.   However the gluino limit still plays an important role in restricting the allowed parameter space for higher values of $s$. The allowed masses of the Higgs and the tentative signal seem to be more compatible with a heavier $Z^\prime$, and also provide information about the cE$_6$SSM parameters even well above the limits from gluinos/squarks and $Z^\prime$.
\subsubsection{Spectrum Generator}
The c$E_6$SSM mass spectrum is calculated from the input parameters using a spectrum generator first written for \cite{Athron:2009bs} where the procedure is described in detail.  However here we summarise the procedure for the purposes of completeness.  The fundamental parameters of the cE$_6$SSM are a unified gauge coupling, Yukawa couplings of the observed fermions, new exotic Yukawa couplings $\lambda_i$ and $\kappa_i$ (where $i = 1..3$) and universal soft masses $m_0$, $M_{1/2}$ and $A$.  The gauge and Yukawa couplings  and the combination of VEVs $v^2 = v_1^2 + v_2^2$ are constrained by experiment at low energies.

To find solutions consistent with both the high scale universality constraints and this low energy data we evolve between the high and low scales with RGEs\footnote{As already described in section \ref{cE6SSM}, we used two loop RGEs for gauge couplings, Yukawa couplings, gaugino masses and trilinear soft couplings and one loop RGEs for scalar masses.}  and impose EWSB conditions by finding simultaneous solutions to the three quadratic minimisation equations which at tree level are functions of $\{ m_0, M_{1/2}, A, s, v_u, v_d, \lambda(Q), g, g^\prime, g_1^\prime \}$.  Since $g$ and $g^\prime$ are fixed by experiment and $g_1^\prime$ is then fixed from requiring gauge coupling unification of all four gauge couplings the three constraints mean that we must have three of  $\{ m_0, M_{1/2}, A, s, \tan \beta, \lambda(Q)\}$ as outputs to fix $M_Z = 91.1876$ GeV, and we chose $m_0$, $M_{1/2}$ and $A$ since the values of $\tan \beta$ and $\lambda(Q)$ are required at the very outset of the calculation to perform the SUSY RG evolution.

Since the three quadratic equations correspond to one quartic equation we have 4 solutions for the soft masses for each set of input parameters,  so our procedure yields between zero and four real solutions for each point.   Finally leading one loop contributions to the effective potential are then included iteratively and the spectrum is then determined.

   Therefore our input parameters are $\{ \lambda_i(M_X), \kappa_i(M_X), s, \tan \beta \}.$  Since we are interested in constraints coming from squarks and gluinos in the parameter space scans we fix $\kappa_{1,2,3} = \kappa$ to keep the colored exotics heavy but since the inert Higgsinos and inert Higgs are only weakly produced we left $\lambda_{1,2} = 0.1$ which is consistent with the previous study of the parameter space where we focussed on lighter scenarios \cite{Athron:2009bs}.  So here we investigate a subspace of the full model where we have only four free parameters $\lambda = \lambda_3, \kappa, \tan \beta, s$.

Having the soft masses as output parameters makes finding iterative solutions including leading tadpole terms trickier than in the MSSM where $\tan\beta$ and $M_Z$ can be traded for $\mu$ and $B\mu$. Without knowing the soft masses at the outset the stop masses cannot be estimated initially and starting without the leading one loop contributions in the EWSB iteration can lead to a significant border region in the parameter space where an EWSB condition can be found for the one loop effective potential but missed in the tree level approximation.

 To resolve this we perform the parameter space scans over a fine grid (rather than employing more sophisticated random sampling scans) and use the solution from the previous step as an initial guess for the tadpoles in the next step. This approach leads to $m_0-M_{1/2}$ plots where the density of the points varies so some regions are very densely populated, while others are sparsely populated. In addition the lower stability of a routine outputting soft masses from the EWSB conditions and the fact that we have a multiplicity of solutions for the soft masses, leading to folds in the parameter space where obtaining a solution can be dependent on the direction one moves through it, renders the search for solutions a non-trivial task even with a fine grid scan.  Due to these issues we also cannot guarantee that we find all potential solutions, however it is clear that certain regions must be excluded for reasons stated in the subsequent text.

The spectrum is then calculated using the expressions presented in \cite{Athron:2009bs}.  The most important constraints come from the gluino and the Higgs, so we include one loop shifts to pole for gluino mass and leading two loop corrections for the lightest Higgs mass.
\subsubsection{Higgs and Gluino contours}
In each plot we fix $s$ and $\tan \beta$ then vary $\lambda$ between $-3$ and $0$ (thus fixing $\mu_{\rm eff} < 0$) and $\kappa$ between $0$ and $3$.  The restriction to $\mu_{\rm eff} < 0$ is to remove a confusing bifurcation in the gluino contours at large $m_0$.

\begin{figure}[th]
\begin{tabular}{cc}
\resizebox{!}{12cm}
{\includegraphics{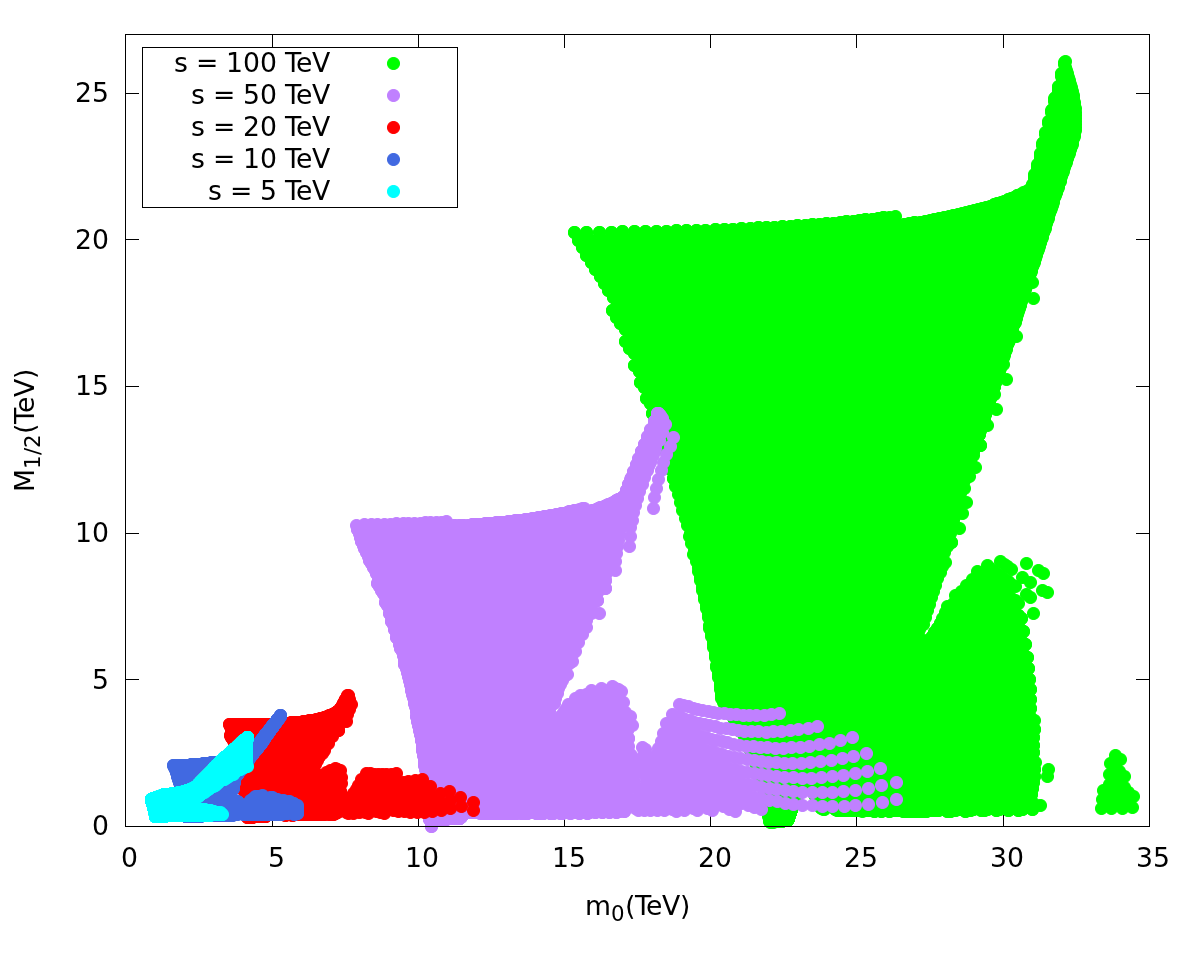}} &
%%\resizebox{!}{6cm}{\includegraphics{Bench_A2_Big_newNotation.png}} \\
\end{tabular}
\caption{Allowed region for cE$_6$SSM with $\tan\beta = 10$, $s =5, 10,20, 50, 100$ TeV in cyan, blue, red, purple and  green respectively with  $\lambda_{12} = 0.1$, while contours are produced with a universal $\kappa$ coupling and $\lambda_3$ are varied, with $\lambda_3$ (and hence $\mu_{\rm eff}$) $\leq 0$.   \label{Fixeds-Allowed} }
\end{figure}

First in light of the exciting progress and effectiveness of the LHC in probing scenarios with multiple TeV scenarios we update Fig.~4 from \cite{Athron:2009bs} to show the full range of $m_0-M_{1/2}$ rather than cutting off at around a TeV and also show even very heavy values of $s$ which are not immune to LHC searches and would will certainly be probed by a Higgs mass measurement. In Fig.~\ref{Fixeds-Allowed} we plot the allowed values of $m_0-M_{1/2}$ for all the solutions we find for $\tan \beta = 10$ and fixed values of $s = 5, 10, 20, 50, 100$ TeV. Since here we are going to use the plots to help us estimate the LHC constraints in the cE$_6$SSM and also to match the plot we are updating, we impose only constraints prior to the LHC, which are specified in \cite{Athron:2009bs}.  Also note in order to match more closely results published by the experimental analyses (where they interpret the search constraints in terms of the cMSSM) we have switched the $x$ and $y$ axes so that $M_{1/2}$ now appears on the y-axis and $m_0$ on the x-axis. 

 Notice that while increasing $s$ pushes up the lower limit on $m_0$ (below this the inert Higgs run below their LEP limit and rapidly become tachyonic), and the upper limit on $M_{1/2}$ (above which no solution satisfying both EWSB and universality conditions can be found) also increases, low $M_{1/2}$ is always possible and is only bounded by the constraints from gauginos (the update of which we will discuss shortly).  Therefore squark/gluino searches can always impact part of the parameter space for any value of $s$ albeit an increasing small fraction of the total parameter space as we go up in $s$.

Note also that in addition to some regions being sparsely populated, the reasons for which are explained above, we also see a significant gap in the plot of allowed solutions for each value of $s$.  Interestingly the solutions for $\mu_{\rm eff} > 0$ (not shown here) although covering a substantially smaller region of the parameter space do tend to cover these gap regions.

Now we want to understand how the squark/gluino searches and Higgs limits constrain the parameter space and where (or if) we can fit the tentative Higgs signal.

\begin{figure}[th]
\begin{tabular}{cc}
\resizebox{!}{6.3cm}
{\includegraphics[trim = 1mm 1mm 1mm 1mm, clip]{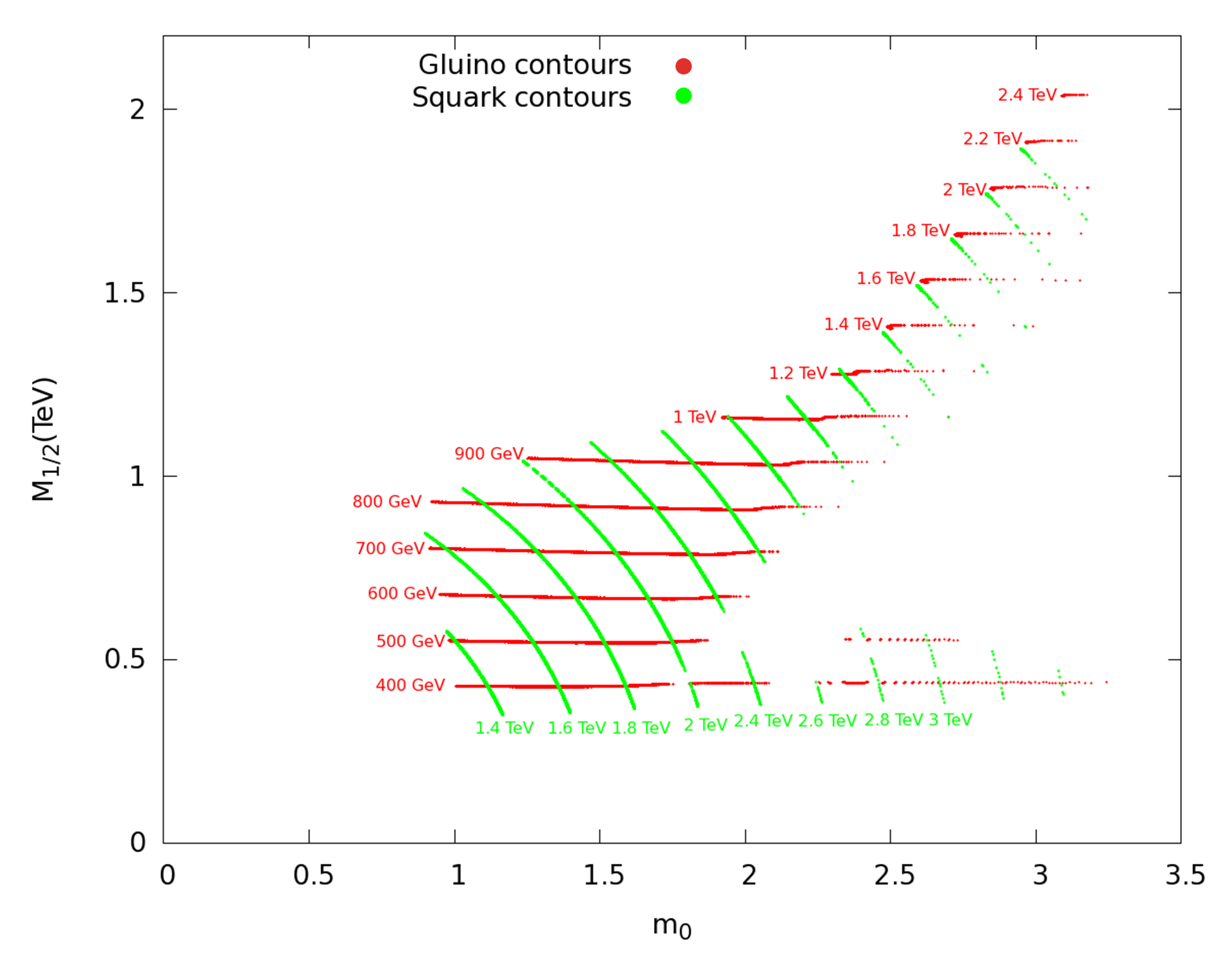}}
%%\resizebox{!}{6cm}{\includegraphics{Bench_A2_Big_newNotation.png}} \\
\resizebox{!}{6.3cm}
{\includegraphics[trim = 1mm 1mm 1mm 1mm, clip]{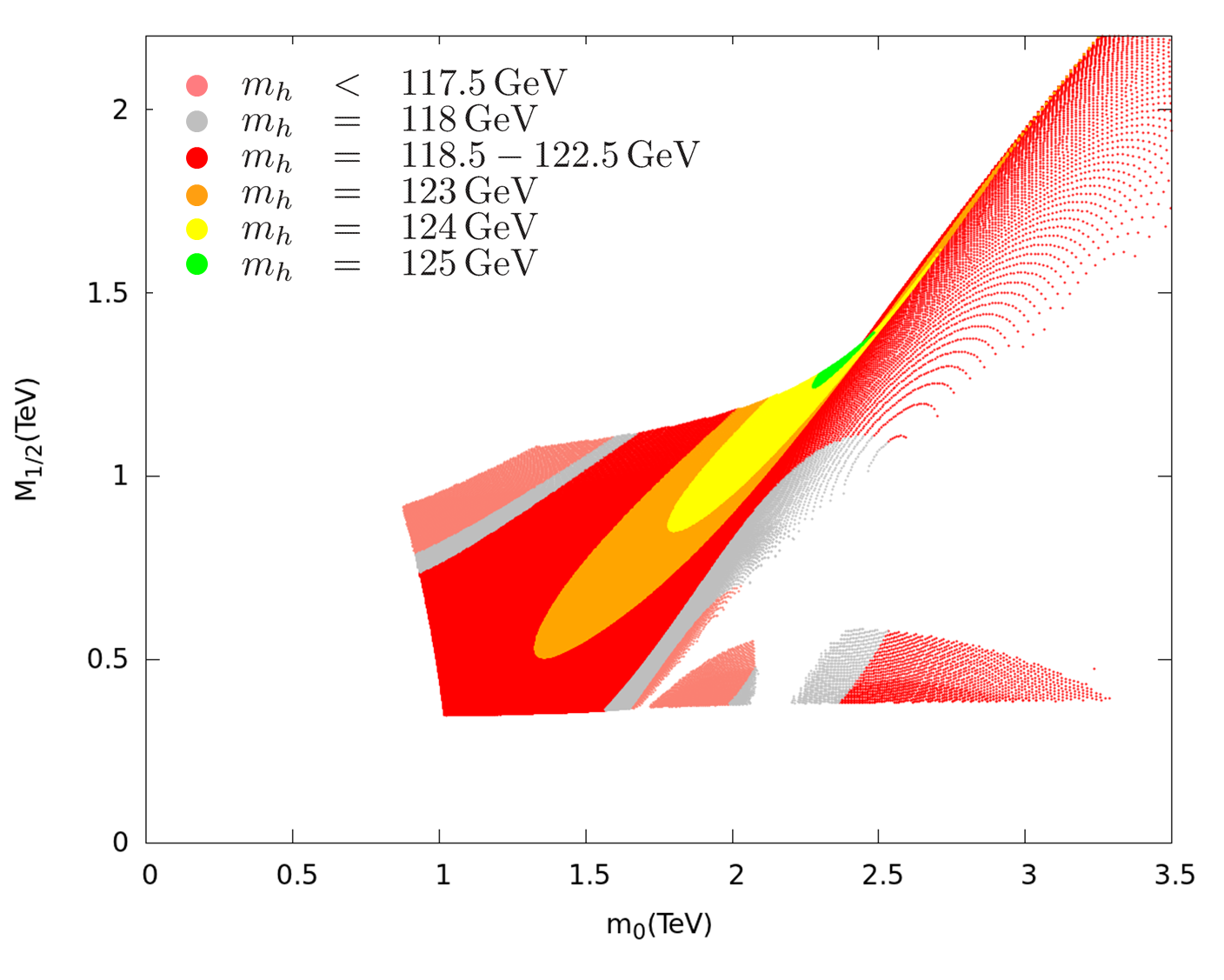}}
\end{tabular}
\caption{Squark and gluino mass contours (left panel) and Higgs mass contours
(right panel) in the
$(m_0,M_{1/2})$ plane of the cE6SSM with with $\tan\beta = 10$, $\lambda_{12} = 0.1$, $s = 5$ TeV, corresponding to $M_{Z^\prime} =  1.889$ TeV. Scans are produced with a universal $\kappa$ coupling varied over $\{ 0,3 \}$ and $\lambda_3$ over $\{ -3,0 \}$  respectively so that $\mu_{\rm eff} \leq 0$.   \label{tb10s5} }
\end{figure}

In Fig.~\ref{tb10s5}, the squark and gluino contours are shown (left) beside the Higgs masses (right) for $\tan \beta = 10, s = 5$ TeV. The squark contours are specifically of the left handed squark mass for the first two generations (it is contours of this squark mass which were plotted on the ATLAS and CMS papers). Both those and the gluino contours are formed by selecting points from the scans where the mass lies in a suitably narrow range such that the width of the contour gives a resolvable line for the scale of the plot.

 For any given gluino mass we find that squark masses must be substantially heavy.  For example if we take the latest limits on the gluino coming from the LHC which apply in the large $m_0$ region, which is about $840$ GeV as described in section \ref{sparticle_searches}, then we find no solutions with squarks below $1.8$ TeV, rendering the larger limits found for the cMSSM (see Fig.~7a of Ref.~\cite{ATLAS_susy2}) in the low $m_0$ region irrelevant.
Therefore we conclude that the squark and gluino constraints place a limit on $M_{1/2}$ at around $1$ TeV.

The right plot showing Higgs masses gives a very different picture.  The Higgs mass varies over the plane (driven to a substantial degree by the variation of $\lambda$) in a very non trivial manner. Different values of the Higgs mass are plotted where the values quoted in the key are the central values in bin with a $\pm 0.5$ GeV width. Remarkably the Higgs mass varies over the plane within a very narrow set of values that include the mass of the tentative signal.

  A substantial region of the parameter space is ruled out by having $m_h$ in the range $118.5-122.5$ GeV (red) and $m_h < 117.5$ GeV (pink). However, in addition to the $118$ GeV (grey) gap in the lower mass exclusion, there is still a significant region towards the centre of the plot which is allowed and even where the tentative signal can be matched. This is mainly for Higgs masses of 124 GeV, as 125 GeV only gives a small region and a $126$ GeV Higgs cannot be realised for these choices of $s$ and $\tan \beta$. If we also take into account the new LHC gluino constraints one can see that little of the $124$ GeV signal is affected while most of the available space for a $123$ GeV  Higgs is removed.

However there are still strong constraints on the parameter space coming from the Higgs limits alone, and it is clear one must be careful to apply both of these complimentary constraints in order to understand where the viable regions are.

So while $s= 5$ TeV has a substantial portion of allowed parameter region which can accommodate the tentative Higgs signal, a measurement of the Higgs mass of 126 GeV, could potentially rule this out.  Therefore it is important to consider other slices of our four dimensional parameter space.

\begin{figure}[t]
\begin{tabular}{cc}
\resizebox{!}{6.4cm}
{\includegraphics[trim = 1mm 1mm 1cm 1mm, clip]{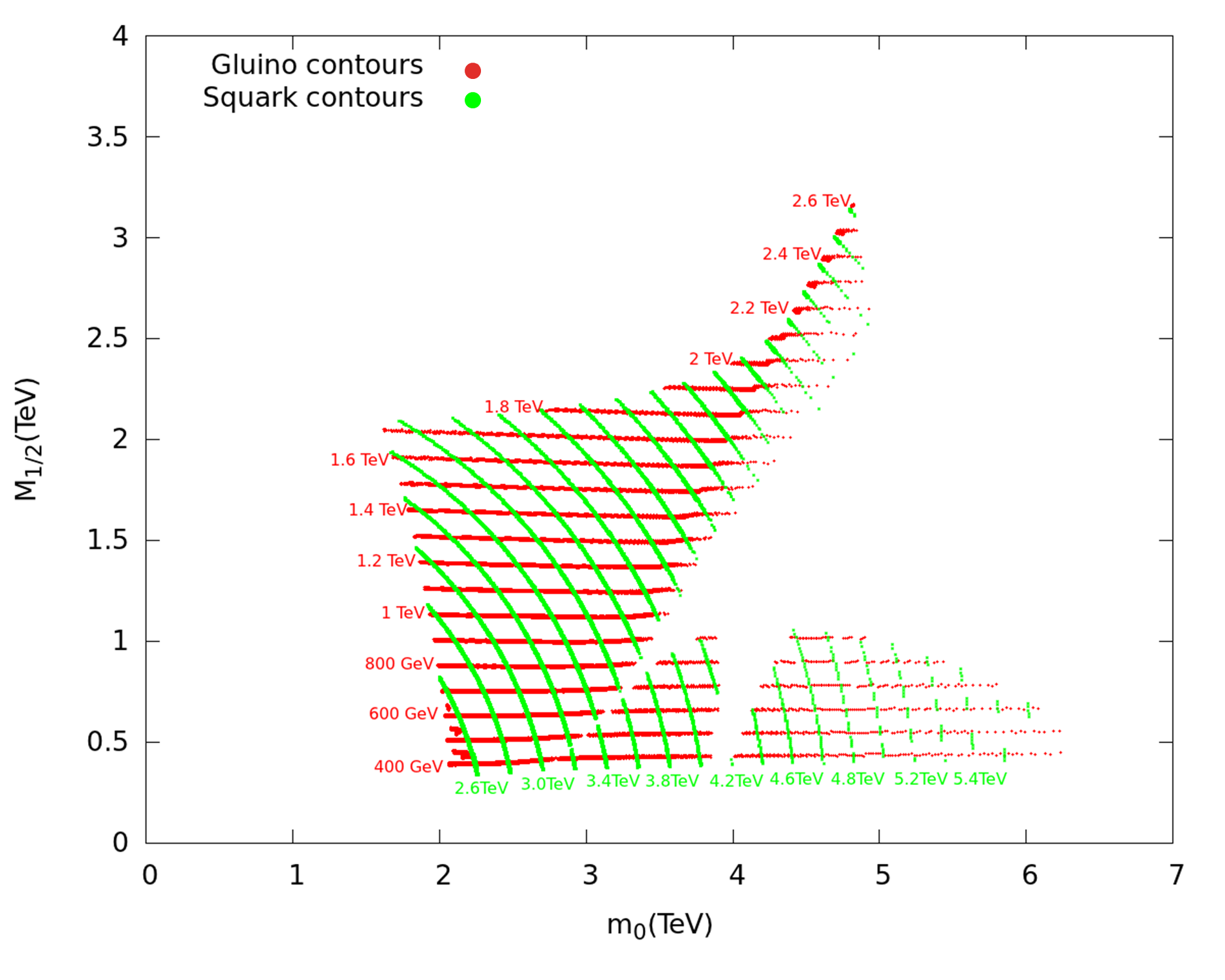}}
%%\resizebox{!}{6cm}{\includegraphicsBe{nch_A2_Big_newNotation.png}} \\
\resizebox{!}{6.4cm}
{\includegraphics[trim = 1mm 1mm 1cm 1mm, clip]{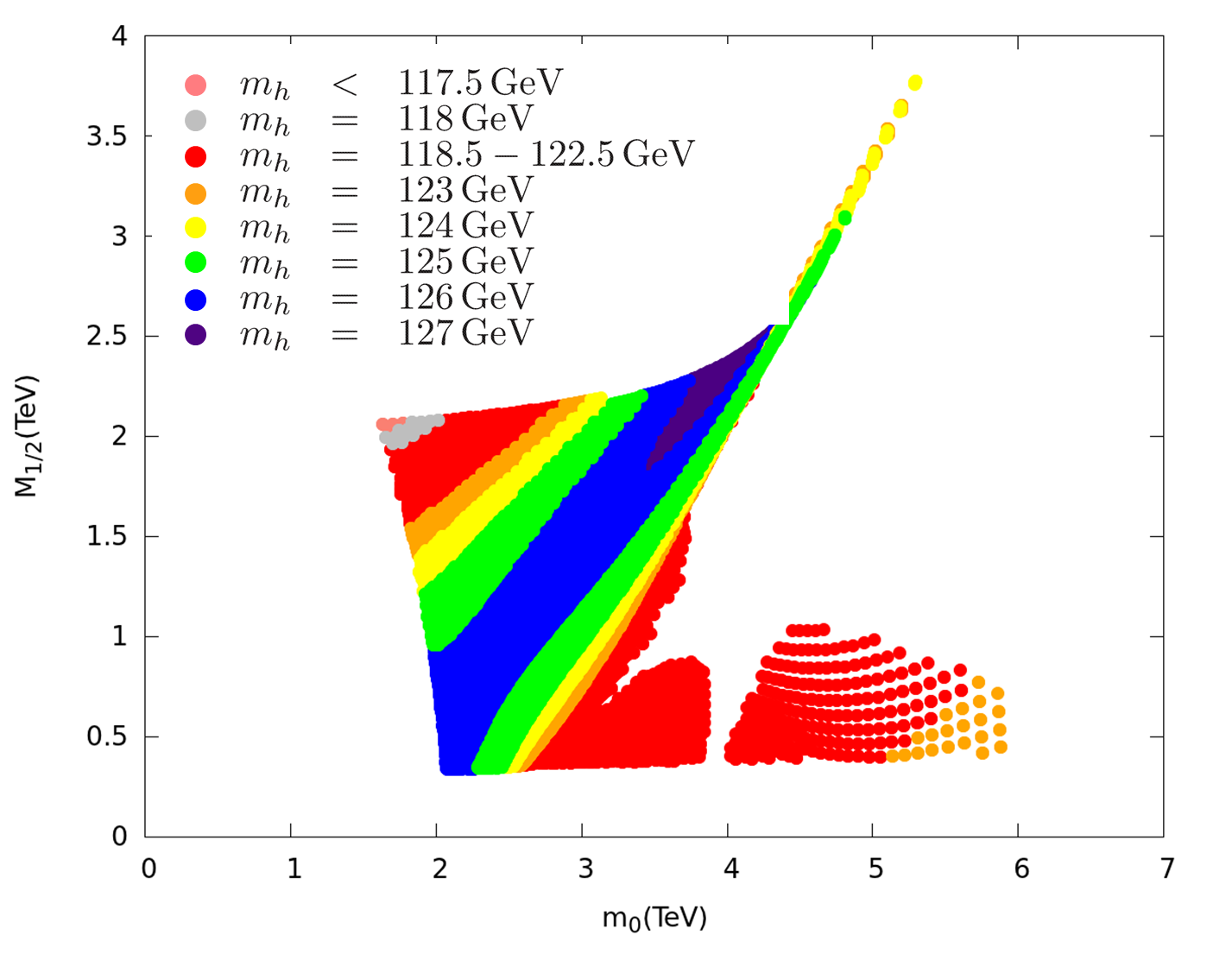}}
\end{tabular}
\caption{Squark and gluino mass contours (left panel) and Higgs mass contours
(right panel) in the $(m_0,M_{1/2})$ plane of the cE6SSM with $\tan\beta = 10$, $\lambda_{12} = 0.1$, $s = 10$ TeV, corresponding to $M_{Z^\prime} =  3.778$ TeV. Scans are produced with a universal $\kappa$ coupling varied over $\{ 0,3 \}$ and $\lambda_3$ over $\{ -3,0 \}$  so that $\mu_{\rm eff} \leq 0$.   \label{tb10s10} }
\end{figure}

In Fig.~\ref{tb10s10} we increase $s$ to $10$ TeV, but keep $\tan \beta =10$ and from the left plot we can see that again the squark and gluino search constraints can all be satisfied by a simple cut on $M_{1/2}$ at $1$ TeV, but now there is a substantially greater proportion of the parameter space which is above this limit since the upper limit on $M_{1/2}$ is increasing with $s$.  Additionally, in the right plot, we also see that all masses of the tentative Higgs signal region can be  comfortably accommodated and this region fills most of allowed the parameter space that is above the $M_{1/2} > 1$ TeV limit we have set from squark and gluino searches.

\begin{figure}[p]
\begin{tabular}{cc}
\resizebox{!}{6.4cm}
{\includegraphics[trim = 1mm 1mm 1cm 1mm, clip]{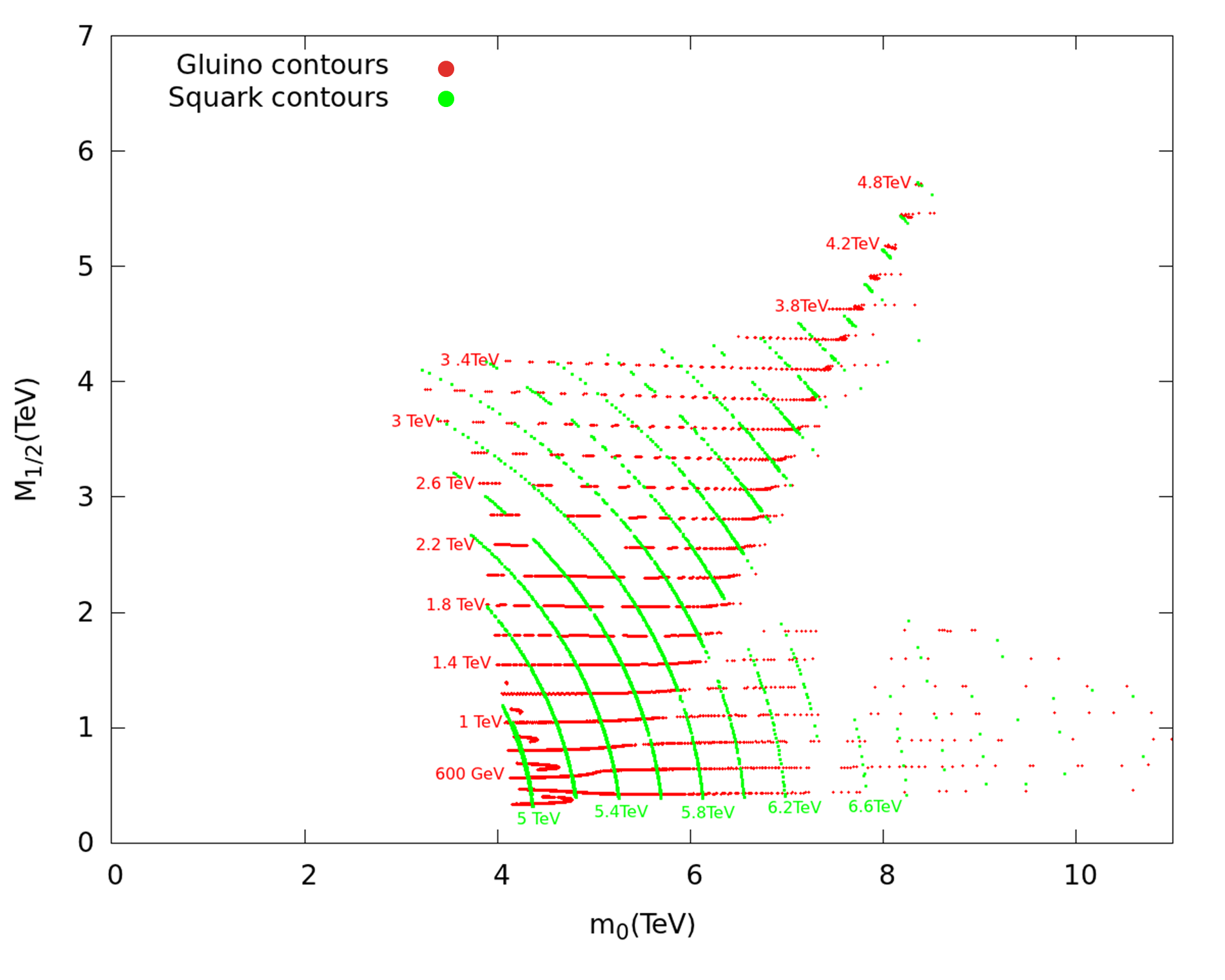}}
\resizebox{!}{6.4cm}
{\includegraphics[trim = 1mm 1mm 1cm 1mm, clip]{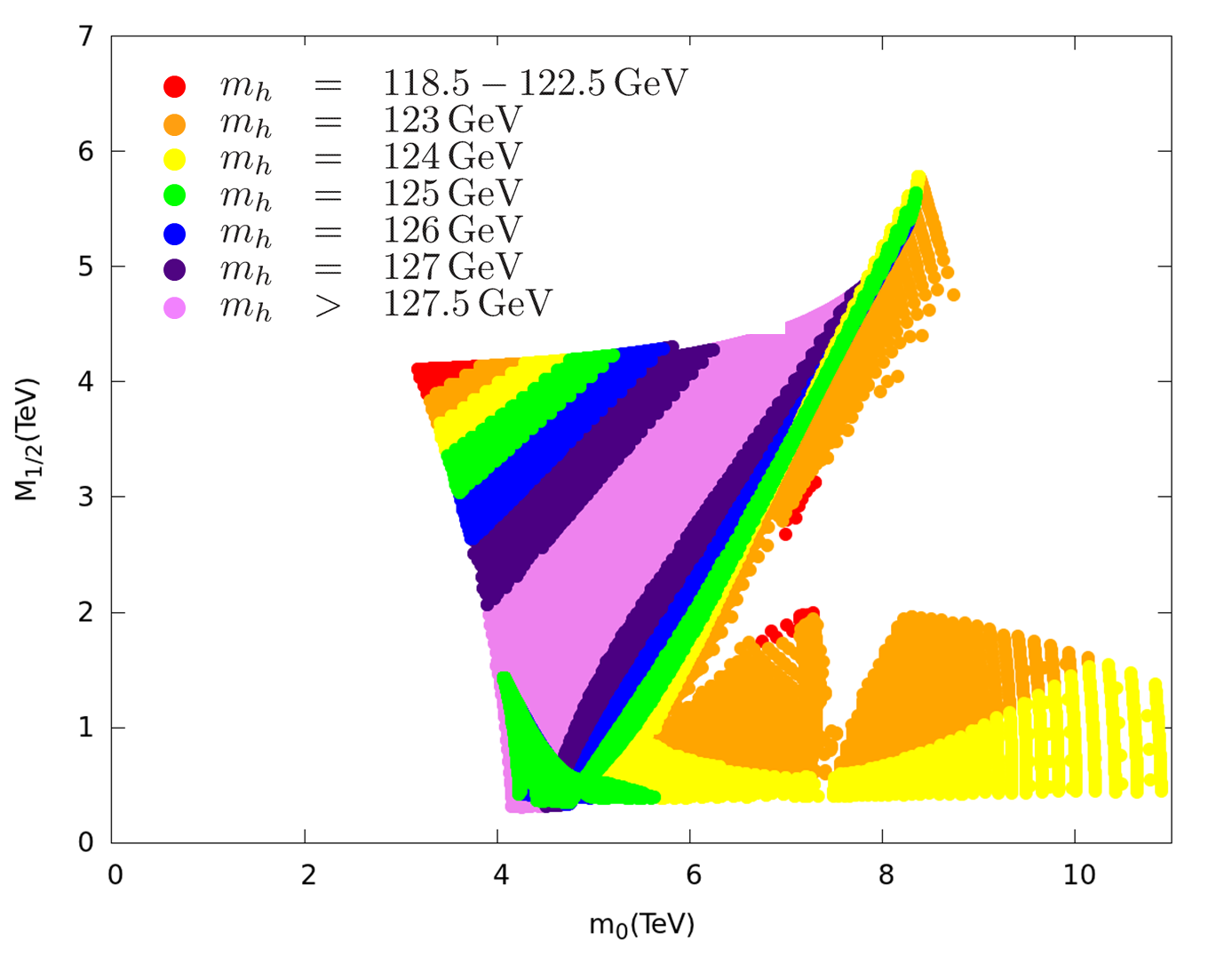}}
\end{tabular}
\caption{Squark and gluino mass contours (left panel) and Higgs mass contours
(right panel) in the $(m_0,M_{1/2})$ plane of the cE6SSM with $\tan\beta = 10$, $\lambda_{12} = 0.1$, $s = 20$ TeV, corresponding to $M_{Z^\prime} =  7.564$ TeV. Scans are produced with a universal $\kappa$ coupling varied over $\{ 0,3 \}$ and $\lambda_3$ over $\{ -3,0 \}$  so that $\mu_{\rm eff} \leq 0$.   \label{tb10s20} }
\end{figure}

In Fig.~\ref{tb10s20} we see that if we increase the singlet VEV further to $s= 20$ TeV then we are no longer restricted by the lower limits on the Higgs mass, with only a few points having a Higgs mass of $122$ GeV, but now there is a substantial region ruled out by the upper limit $m_h \geq 127.5$ set by CMS.

\begin{figure}[p]
\begin{tabular}{cc}
\resizebox{!}{6.4cm}
{\includegraphics[trim = 1mm 1mm 1cm 1mm, clip]{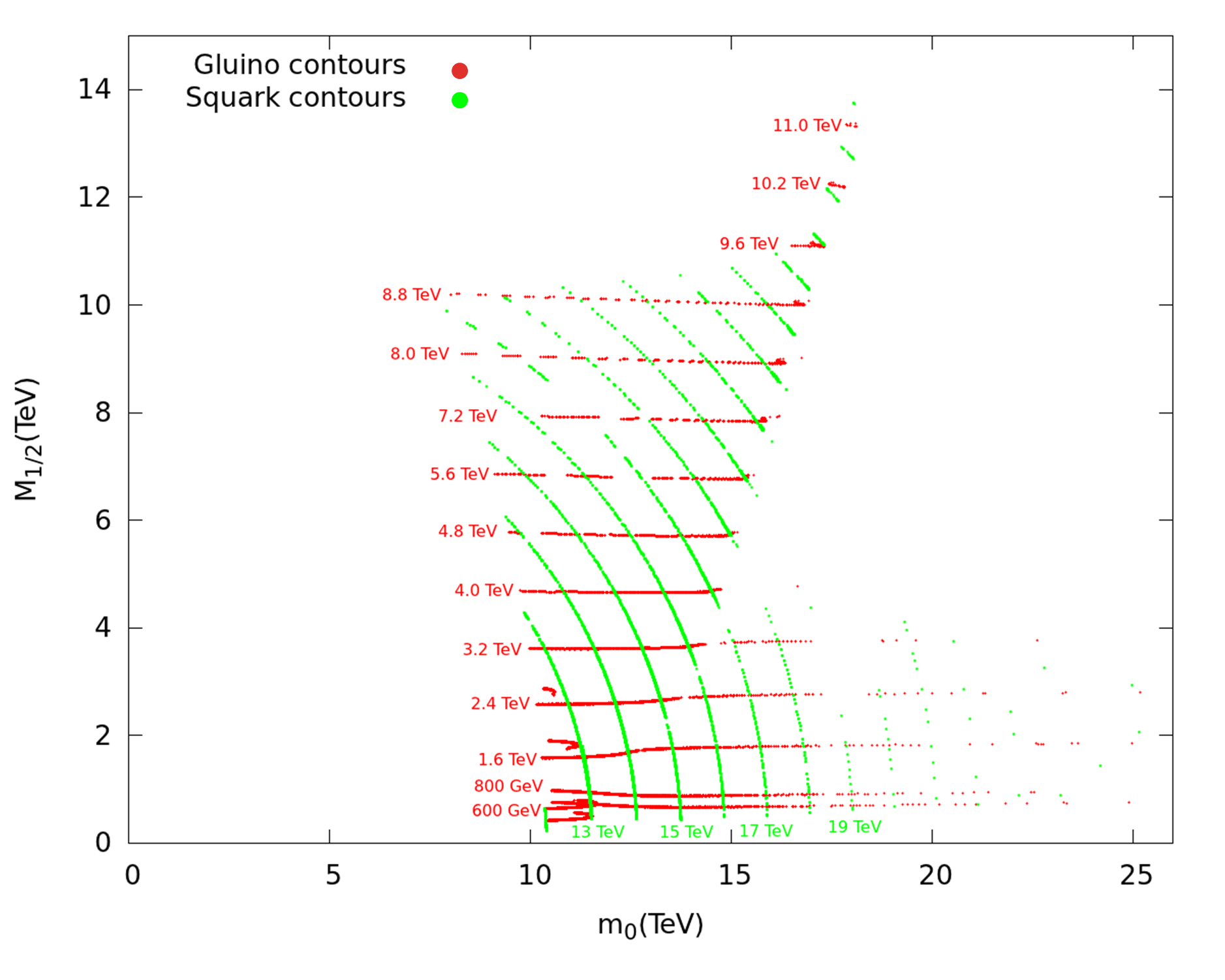}}
\resizebox{!}{6.4cm}
{\includegraphics[trim = 1mm 1mm 1cm 1mm,  clip]{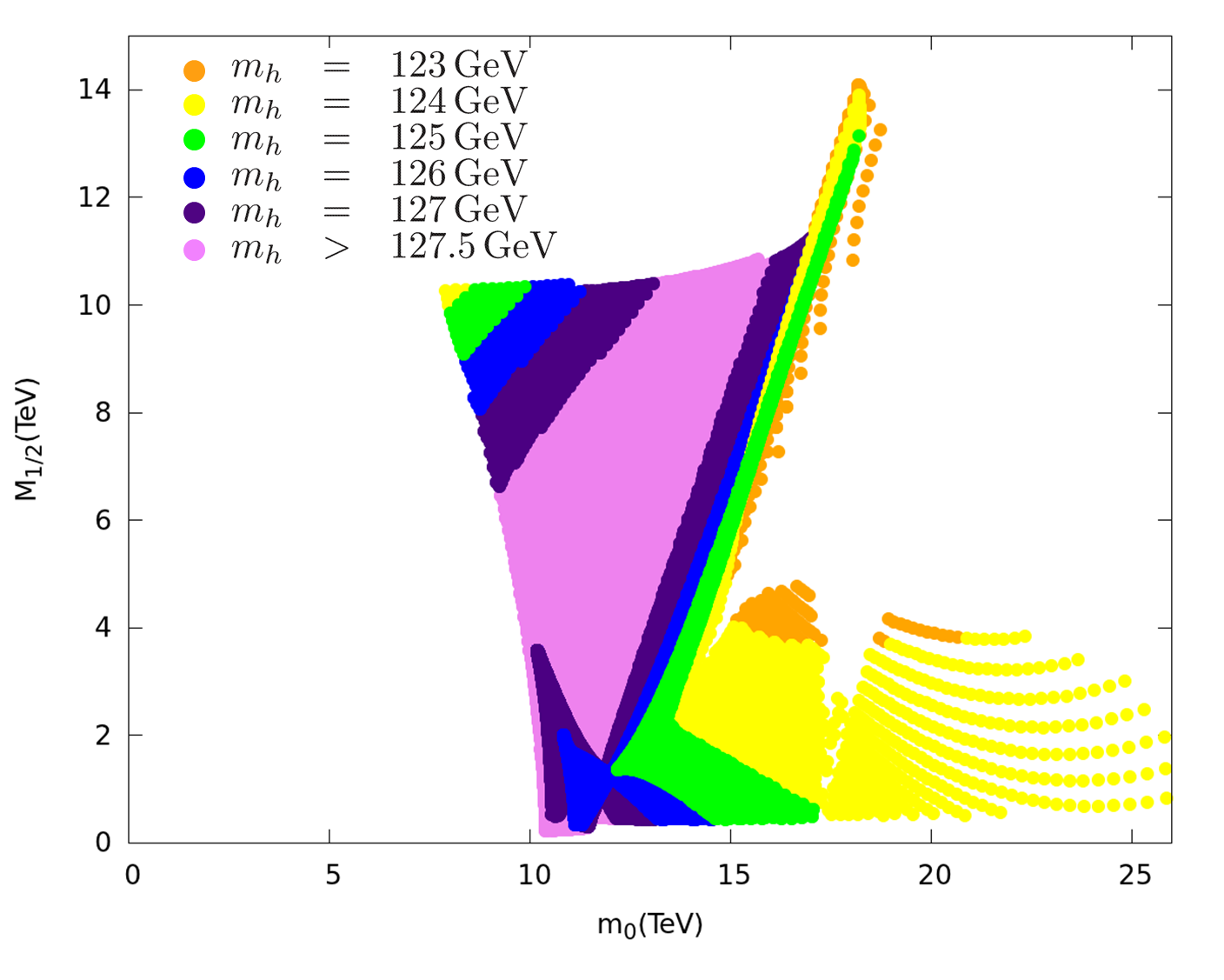}}
\end{tabular}
\caption{Squark and gluino mass contours (left panel) and Higgs mass contours
(right panel) in the $(m_0,M_{1/2})$ plane of the cE6SSM with  $\tan\beta = 10$, $\lambda_{12} = 0.1$, $s = 50$ TeV, corresponding to $M_{Z^\prime} =  18.996$ TeV. Scans are produced with a universal $\kappa$ coupling varied over $\{ 0,3 \}$ and $\lambda_3$ over $\{ -3,0 \}$  so that $\mu_{\rm eff} \leq 0$.   \label{tb10s50} }
\end{figure}

\begin{figure}[tp]
\begin{tabular}{cc}
\resizebox{!}{6.3cm}
{\includegraphics[trim = 1mm 1mm 9mm 1mm, clip]{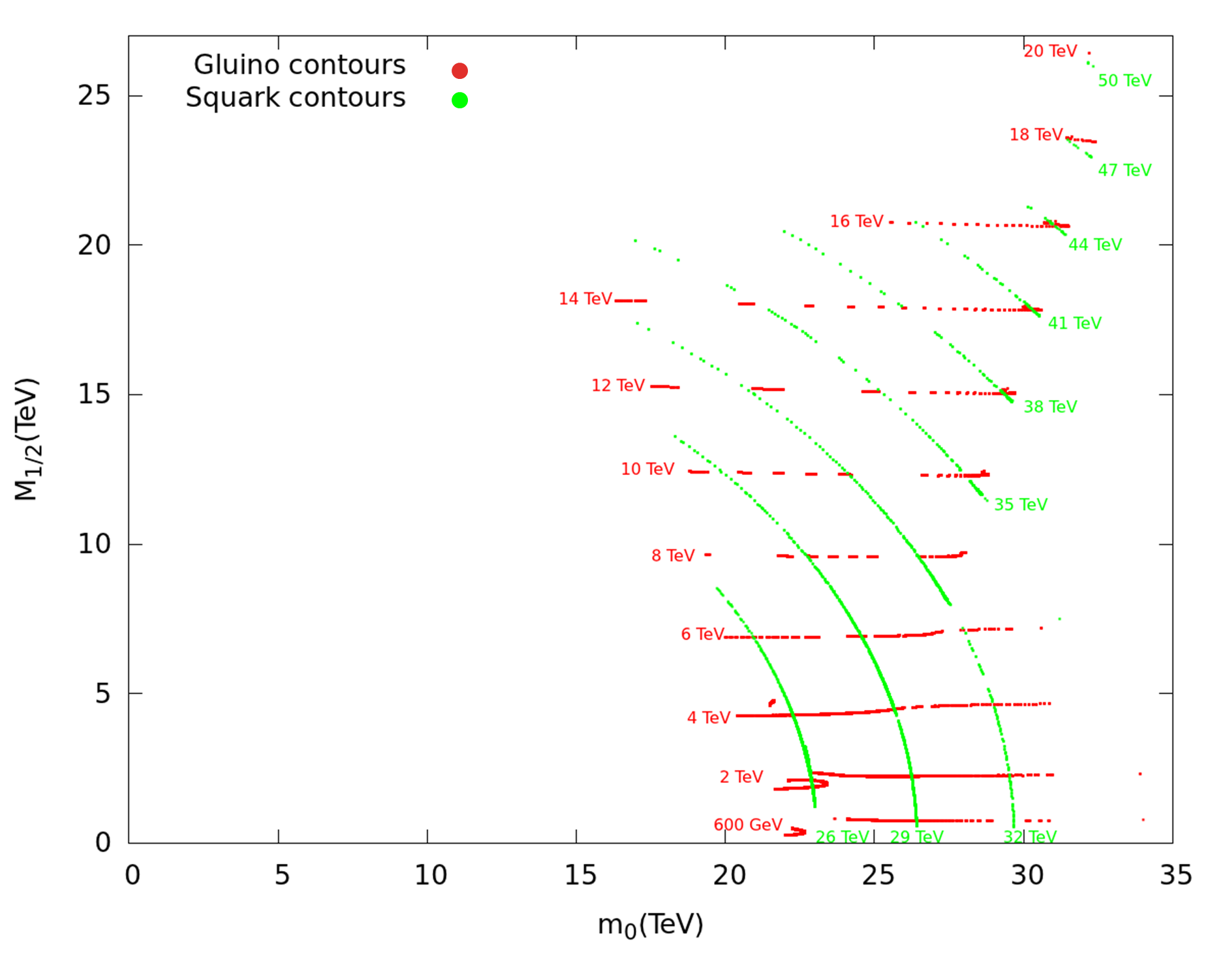}}
\resizebox{!}{6.3cm}
{\includegraphics[trim = 1mm 1mm 9mm 1mm, clip]{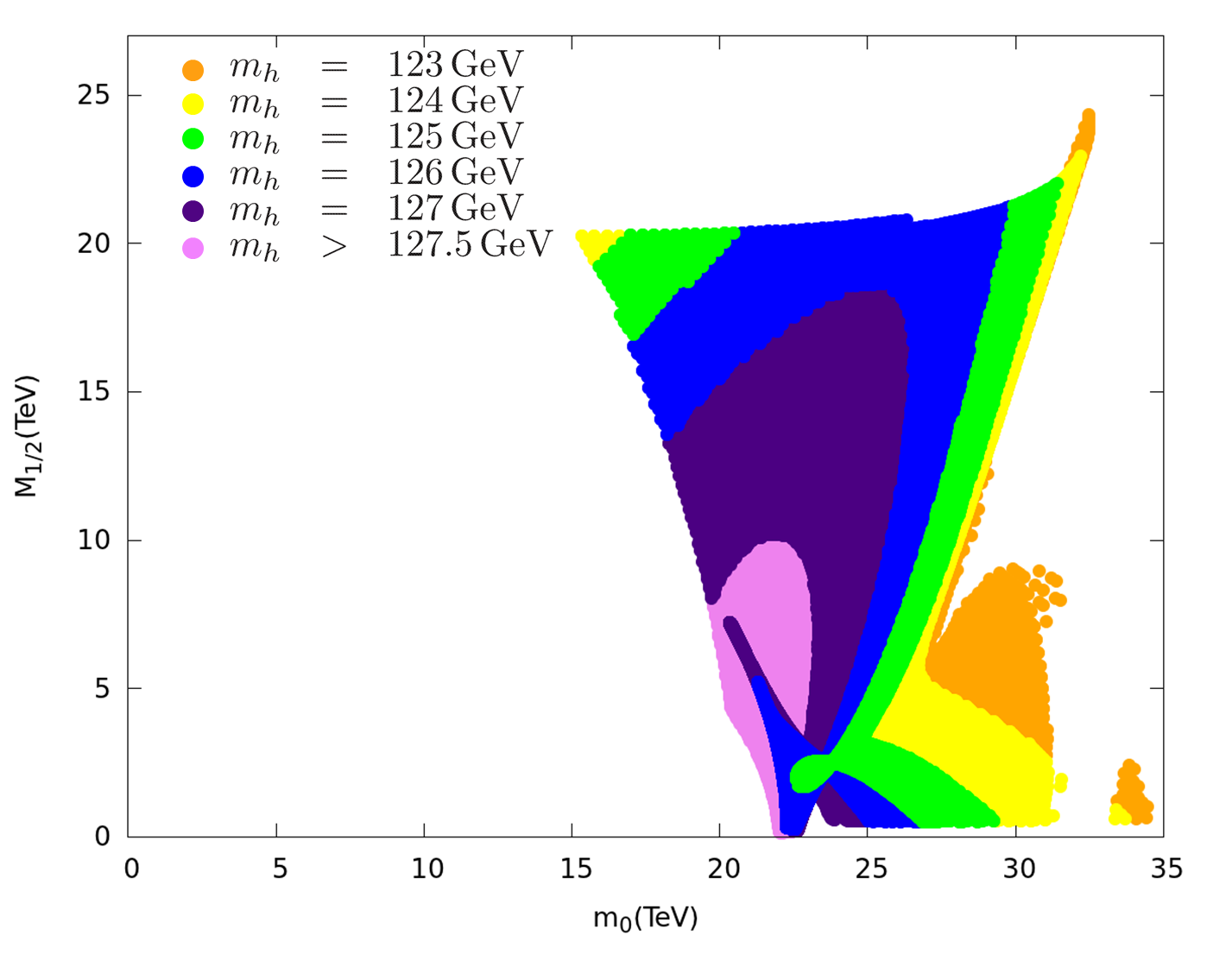}}
\end{tabular}
\caption{Squark and gluino mass contours (left panel) and Higgs mass contours
(right panel) in the $(m_0,M_{1/2})$ plane of the cE6SSM with$\tan\beta = 10$, $\lambda_{12} = 0.1$, $s = 100$ TeV, corresponding to $M_{Z^\prime} =  37.808$ TeV. Scans are produced with a universal $\kappa$ coupling  varied over $\{ 0,3 \}$ and $\lambda_3$ over $\{ -3,0 \}$ so that $\mu_{\rm eff} \leq 0$.   \label{tb10s100} }
\end{figure}

Figs.~\ref{tb10s50} and \ref{tb10s100} demonstrate that even with very heavy $s$ values, such that the $Z^\prime$ is well beyond reach of the LHC, not only is there still a small region of parameter space where the gluino is observable, but additionally a Higgs mass measurement would yield useful information about the parameter space well above what can actually be constrained from direct searches. This illustrates the significance of the Higgs to providing constraints and measurement of cE$_6$SSM parameters.

 Notice also that while in much of the parameter space new physics states are out of reach, reducing the $\lambda_{1,2}$ coupling such that the inert Higgsinos are observable would not perturb the RG evolution much, so these plots remain a very good approximation.  Thus they reveal an interesting potential scenario where only the inert Higgsinos and the SM-like Higgs are discovered, but an accurate Higgs mass measurement would give a great deal of information on the parameter space.

\begin{figure}[th]
\begin{tabular}{cc}
\resizebox{!}{6.3cm}
{\includegraphics[trim = 1mm 1mm 1cm 1mm, clip]{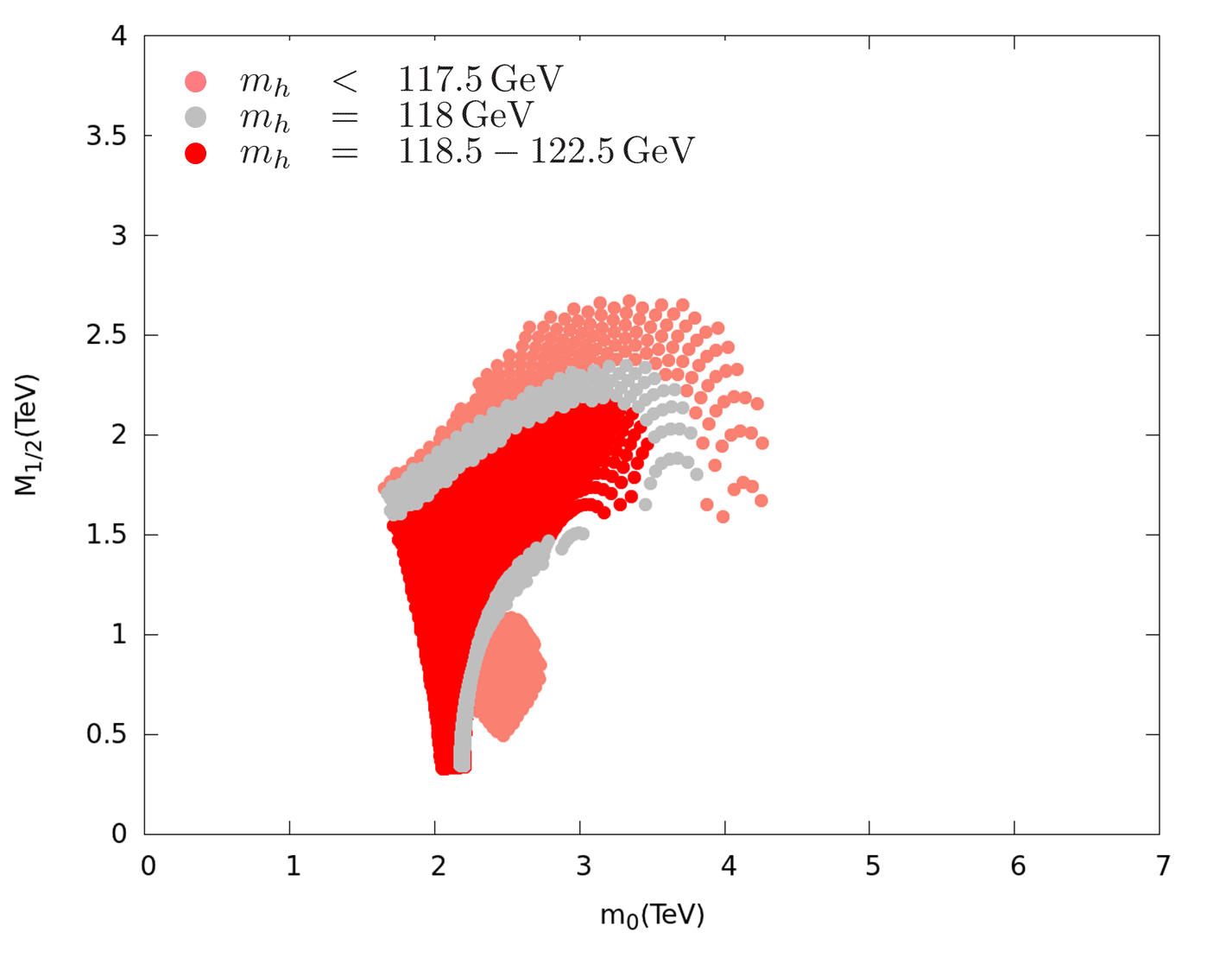}}
\resizebox{!}{6.3cm}
{\includegraphics[trim = 1mm 1mm 1cm 1mm, clip]{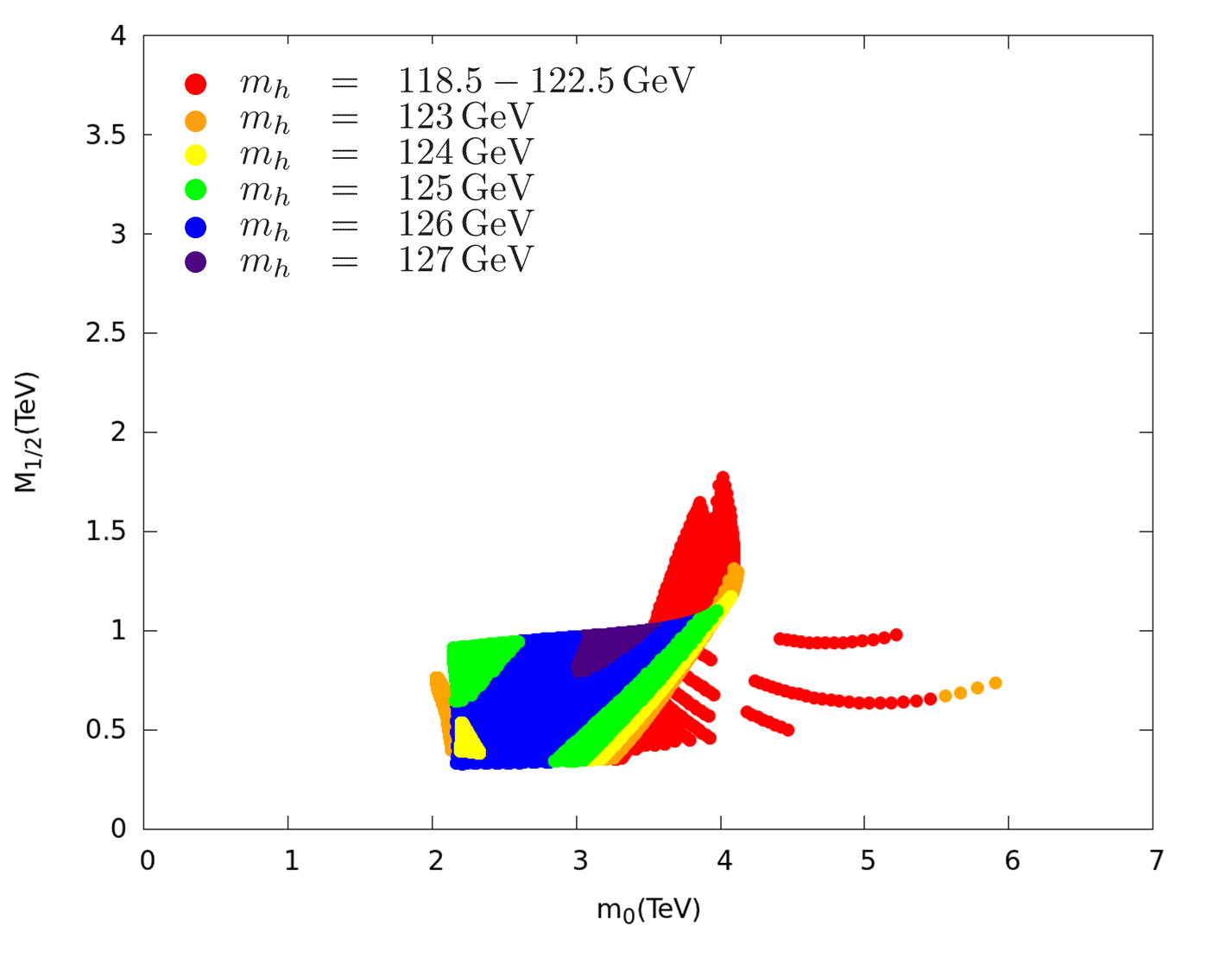}}
\end{tabular}
\caption{Higgs mass contours for $\tan\beta = 3$ (left panel) and $\tan\beta = 30$ (right panel) in the $(m_0,M_{1/2})$ plane of the cE6SSM with $\lambda_{12} = 0.1$, $s = 10$ TeV, corresponding to $M_{Z^\prime} =  3.779$ TeV. Scans are produced with a universal $\kappa$ coupling varied over $\{ 0,3 \}$ and $\lambda_3$ over $\{ -3,0 \}$  so that $\mu_{\rm eff} \leq 0$.   \label{s10tb3-30} }
\end{figure}

Finally we comment on the $\tan \beta$ dependence of these results.  The form of the squark and gluino contours is not substantially modified by changing $\tan \beta$ so we do not reproduce these plots here. However the allowed region of parameter space is dramatically changed, as are the Higgs masses.  This is illustrated in Fig.~\ref{s10tb3-30} where we plot the allowed region of the parameters space for $s = 10$ TeV and $\tan\beta = 3$ (left) and $\tan\beta = 30$ (right). Here we see that the combination  $s = 10$ TeV and $\tan\beta = 3$ is almost entirely ruled out with only the $118$ GeV window left.   On the other hand for $\tan\beta = 30$ most of the parameter space is compatible with the tentative Higgs signal and, in particular, a Higgs of $126$ GeV appears very typical.  However the overall allowed region of the parameter space has significantly shrunk in comparison to the $\tan \beta = 10$ case.

\subsection{Benchmark Points}

We have chosen five benchmark points that reproduce a Higgs mass of around $125\,$GeV, in order to demonstrate significant and interesting features that may arise in the cE$_6$SSM. These benchmarks are given in Table~\ref{table:benchmarks}, labelled HBM1 to HBM5 (for Heavy Bench-Mark). All of these benchmarks evade current experimental constraints, but predict new particle states that may be found at the LHC.  They represent a wide selection of different scales for the scalar VEV, with $s$ ranging from $5\,$TeV up to $100\,$TeV. As discussed in the last section,  low $\tan \beta$ has difficulty reproducing a $125\, {\rm GeV}$ Higgs boson, while high $\tan \beta$ suffers from a restricted parameter space due to the requirements of correct EWSB, so for all these benchmarks we adopt a medium value of $\tan \beta=10$.  All benchmarks have a reasonably low value for the mass of the lightest neutralino, which is a consequence of choosing reasonably low values of $M_{1/2}$ thereby ensuring a small bino mass within reach of the LHC (in these scenarios, the lightest neutralino is always predominantly a bino). This also means that the gluino stays reasonably light in all these scenarios too (though still above current LHC exclusion). Furthermore, all five benchmarks conform to the condition of Eq.~(\ref{DM-cond}) where the lightest inert Higgsino is $10\,$GeV heavier than the lightest neutralino, thereby giving a correct Dark Matter relic abundance, as discussed in section~\ref{DM}

\begin{table}[p]
\begin{center}
\begin{tabular}{|c|c|c|c|c|c|c|c|c|}
  \hline                                                             &  \textbf{\footnotesize HBM1}               &  \textbf{\footnotesize HBM2}        &  \textbf{\footnotesize HBM3}                  &  \textbf{\footnotesize HBM4}                              &  \textbf{\footnotesize HBM5}     \\ \hline
  \footnotesize $\tan \beta$                                              &\footnotesize 10                                   &\footnotesize 10        &\footnotesize 10         	                 &\footnotesize 10         	                                 &\footnotesize 10      \\[-1.5mm]
  \footnotesize $\lambda_3(M_X)$                                           &\footnotesize -0.22                     &\footnotesize -0.35              &\footnotesize -0.55        	                &\footnotesize -0.15        	                              &\footnotesize -0.16799     \\[-1.5mm]	
  \footnotesize $\lambda_{2}(M_X)$                                     &\footnotesize 0.1373                              &\footnotesize 0.141               &\footnotesize 0.035                            &\footnotesize 0.12                                    &\footnotesize 0.1427      \\[-1.5mm]
 \footnotesize $\lambda_{1}(M_X)$                                     &\footnotesize 0.0374                             &\footnotesize 0.0299                &\footnotesize 0.0252                           &\footnotesize 0.006                                   &\footnotesize 0.00237          \\[-1.5mm]
	     	
    \footnotesize $\kappa_3(M_X)$                                        &\footnotesize 0.17                       &\footnotesize 0.45               &\footnotesize 0.9    	                     &\footnotesize 0.9    	                                   &\footnotesize 0.3655         \\[-1.5mm]	
\footnotesize $\kappa_{1,2}(M_X)$                                      &\footnotesize 0.17                      &\footnotesize 0.02                  &\footnotesize 0.02        	             &\footnotesize 0.015       	                                   &\footnotesize 0.3655              \\[-1.5mm]	
\footnotesize $s$[TeV]                                                 &\footnotesize 5                       &\footnotesize 10      	               &\footnotesize 20         	               &\footnotesize 50         	                            &\footnotesize 100          \\[-1.5mm]
\footnotesize $M_{1/2}$[GeV]                                           &\footnotesize 1135                       &\footnotesize 1570                 &\footnotesize 1847                             &\footnotesize 1259                                             &\footnotesize 1148                    \\[-1.5mm]	 
\footnotesize $m_0$ [GeV]                                              &\footnotesize 2158                     &\footnotesize 2490 	               &\footnotesize 4698                             &\footnotesize 16106                                          &\footnotesize 27109                 \\[-1.5mm]      	
\footnotesize $A_0$[GeV]                                             &\footnotesize -266                      &\footnotesize 2010   		       &\footnotesize 8759    	                       &\footnotesize -1658    	                                     &\footnotesize -24825           \\           \hline

\footnotesize $m_{\tilde{D}_{1}}(3)$[GeV]                               &\footnotesize 2403                      &\footnotesize 5734    	               & \footnotesize   14343                         & \footnotesize   39783                                       &\footnotesize 48516      	\\[-1.5mm]
\footnotesize $m_{\tilde{D}_{2}}(3)$[GeV]                              &\footnotesize  3315                       &\footnotesize  7961                       & \footnotesize    16658  	                     & \footnotesize    40838  	                              &\footnotesize 53511    \\[-1.5mm]
\footnotesize $\mu_D(3)$[GeV]                                           &\footnotesize  1748                    &\footnotesize 6725                   & \footnotesize  15570                          & \footnotesize  39925                                           &\footnotesize 48820        \\[-1.5mm]	
\footnotesize $m_{\tilde{D}_{1}}(1,2)$[GeV]                             &\footnotesize 2403                      &\footnotesize 2366                     & \footnotesize 3141                            & \footnotesize 12435                                                &\footnotesize 48516           \\[-1.5mm]	
\footnotesize $m_{\tilde{D}_{2}}(1,2)$[GeV]                            &\footnotesize 3314                       &\footnotesize  2899                    & \footnotesize 4268                            & \footnotesize 13869                                              &\footnotesize 53511       \\[-1.5mm]	
\footnotesize $\mu_D(1,2)$[GeV]                                         &\footnotesize  1748                     &\footnotesize  368 		       &  \footnotesize 521    	                       &  \footnotesize 1025    	                                      &\footnotesize 48820          \\           \hline

\footnotesize $|m_{\chi^0_6}|$[GeV]                                     &\footnotesize 1982                      &\footnotesize 3908   	             &\footnotesize 7722                             &\footnotesize 1900                                                &\footnotesize 37877           \\[-1.5mm]

\footnotesize $m_{h_3}\simeq M_{Z'}$[GeV]                               &\footnotesize  1889                      &\footnotesize 3779  	            &\footnotesize 7564       	                    &\footnotesize 18996       	                                       &\footnotesize 37808           \\[-1.5mm]
\footnotesize $|m_{\chi^0_5}|$[GeV]                                     &\footnotesize 1802                      &\footnotesize 3655   		     &\footnotesize 7410      	                     &\footnotesize 18822      	                                        &\footnotesize 37740     \\
\hline
\footnotesize $m_S(1,2)$[GeV]                                          &\footnotesize 2567                       &\footnotesize 3680                &  \footnotesize 7148                           &  \footnotesize 20937                                                  &\footnotesize 38076     \\[-1.5mm]
	
\footnotesize $m_{H_2}(2)$[GeV]                                     &\footnotesize  2163                    &\footnotesize 2463   	              &\footnotesize 3491      	   	              &\footnotesize 14680      	   	                                 &\footnotesize 24028       \\[-1.5mm]

\footnotesize $m_{H_1}(2)$[GeV]                                      &\footnotesize 2084                    &\footnotesize 1834   		        &\footnotesize 2440                             &\footnotesize 12151                                              &\footnotesize 18575       \\[-1.5mm]	

\footnotesize $m_{H_2}(1)$[GeV]                                     &\footnotesize  2092                    &\footnotesize 2060   	              &\footnotesize 3460       	   	      &\footnotesize 13728       	   	                         &\footnotesize 21200      \\[-1.5mm]

\footnotesize $m_{H_1}(1)$[GeV]                                      &\footnotesize 2015                    &\footnotesize 1670   		        &\footnotesize 2452  	                        &\footnotesize 12355  	                                         &\footnotesize 17507     \\[-1.5mm]

\footnotesize $\mu_{\tilde{H}}(2)$[GeV]                              &\footnotesize 680                        &\footnotesize 1120  		              &\footnotesize 427                              &\footnotesize 3813                                        &\footnotesize 9967        \\   [-1.5mm]
\footnotesize $\mu_{\tilde{H}}(1)$[GeV]                              &\footnotesize 187                       &\footnotesize 257  		                &\footnotesize 307    	                        &\footnotesize 192    	                                 &\footnotesize 167        \\
\hline

\footnotesize $m_{\tilde{u}_1}(1,2)$[GeV]                              &\footnotesize 2689                      &\footnotesize 3450                      &\footnotesize 5818     	                 &\footnotesize 17254     	                                        &\footnotesize 29663        \\[-1.5mm]
 	
\footnotesize $m_{\tilde{u}_2}\simeq m_{\tilde{d}_1}(1,2)$[GeV]           &\footnotesize  2743                     &\footnotesize 3531                      &\footnotesize 5885                             &\footnotesize 17264                                               &\footnotesize 29668    \\[-1.5mm]	
\footnotesize $m_{\tilde{d}_2}(1,2)$[GeV]                              &\footnotesize  2749                     &\footnotesize 3644      	              &\footnotesize 6285                             &\footnotesize 18260                                              &\footnotesize 31981           \\[-1.5mm]            	 
\footnotesize $m_{\tilde{e}_1}(1,2,3)$[GeV]                             &\footnotesize 2272                     & \footnotesize 2815                   &\footnotesize 5310            	               &\footnotesize 17190            	                                &\footnotesize 29631  \\[-1.5mm]
\footnotesize $m_{\tilde{e}_2}(1,2,3)$[GeV]                            &\footnotesize  2405                     &\footnotesize 3139                   &\footnotesize 5884       	              &\footnotesize 18204       	                                        &\footnotesize 31956         \\[-1.5mm]	
\footnotesize $m_{\tilde{\tau}_1}$[GeV]                                 &\footnotesize 2254                     &\footnotesize 2788     	             &\footnotesize 5230       	                     &\footnotesize 17020       	                                        &\footnotesize 29373           \\[-1.5mm]
\footnotesize $m_{\tilde{\tau}_2}$[GeV]                                 &\footnotesize 2396                      &\footnotesize 3127                     &\footnotesize 5849	 	                 &\footnotesize 18127	 	                                        &\footnotesize 31837           \\[-1.5mm]	
\footnotesize $m_{\tilde{b}_2}$[GeV]                                    &\footnotesize 2729                    &\footnotesize 3510      	                       &\footnotesize 6201    	                       &\footnotesize 18123   	                                &\footnotesize 31767            \\[-1.5mm]        	
\footnotesize $m_{\tilde{b}_1}$[GeV]                                    &\footnotesize 2370                     &\footnotesize 2979      	               &\footnotesize 4621                             &\footnotesize 14632                                             &\footnotesize 25421              \\[-1.5mm]	
\footnotesize $m_{\tilde{t}_2}$[GeV]                                    &\footnotesize 2381                     &\footnotesize 2994                      &\footnotesize 4634      	                 &\footnotesize 14633     	                                        &\footnotesize 25422         \\[-1.5mm]	
\footnotesize $m_{\tilde{t}_1}$[GeV]                                    &\footnotesize 1877                     &\footnotesize 2220      	             & \footnotesize  2877   	                     & \footnotesize  11607  	                                         &\footnotesize 20632       \\
\hline
                                                                                                     	
\footnotesize $|m_{\chi^0_{3,4}}|\simeq |m_{\chi^{\pm}_2}|$[GeV]             &\footnotesize 867                       &\footnotesize 2281       	             &\footnotesize 4897	                     &\footnotesize 3819	                                         &\footnotesize 9398     \\[-1.5mm]         	 	
\footnotesize $m_{h_2}\simeq m_A \simeq m_{H^{\pm}}$[GeV]                 &\footnotesize 1890                    &\footnotesize 2742       	          &\footnotesize 5254    	                  &\footnotesize 5254    	                                        &\footnotesize 19474        \\[-1.5mm]	
\footnotesize $m_{h_1}$[GeV]                                            &\footnotesize 124                       &\footnotesize 124         	        &\footnotesize 124                              &\footnotesize 125                                                &\footnotesize 125      \\    	
\hline								    				              				                                     	                                             	                                                             		 
\footnotesize $m_{\tilde{g}}$[GeV]                                     &\footnotesize 984                       &\footnotesize 1352       	        &\footnotesize 1659        	                &\footnotesize 1129        	                                  &\footnotesize 1001     \\[-1.5mm]	
\footnotesize $|m_{\chi^{\pm}_1}|\simeq |m_{\chi^0_2}|$[GeV]              &\footnotesize 313                        &\footnotesize 439  	                        &\footnotesize 526                              &\footnotesize 324                                       &\footnotesize 280          \\[-1.5mm]         	
\footnotesize $|m_{\chi^0_1}|$[GeV]                                  &\footnotesize 177                         &\footnotesize 247  		                 &\footnotesize 297                              &\footnotesize 182                                       &\footnotesize 157   \\
\hline
\end{tabular}
\caption{Parameters and masses for
the new heavier benchmarks with Higgs masses in the range of the tentative signal at $m_h = 124-125$ GeV.}
\label{table:benchmarks}
\end{center}
\end{table}

HBM1 is an example benchmark with $s=5\,$TeV and a Higgs mass of $124\,$GeV. We have lifted the previous degeneracy $\lambda_1=\lambda_2$ used in our scans over the parameter space, in order to ensure that Eq.~(\ref{DM-cond}) is satisfied. However, since varying $\lambda_2$ will only effect the mass of the inert Higgs and Higgsinos, this can be done for any point that we found in our scans, yielding an identical spectrum except for the inerts. Consequently, HBM1 may be thought of as one of the points seen in the yellow region of Fig.~\ref{tb10s5}.  Since this is our benchmark with the lowest value of $s$, it also contains the lightest $Z^\prime$ with a mass of $1889\,$GeV, just a little beyond current LHC bounds and possibly detectable reasonably soon. The rather small value of $m_0$ results in reasonably light squarks and sleptons that would be observable at the LHC once more luminosity is gathered. Finally, since $\kappa_1=\kappa_2=\kappa_3$, the scalars $\tilde D_1$ and $\tilde D_2$ are separately degenerate over the three generations, and are light enough to be produced at the LHC. Recall that these scalars are even under the analogue of R-parity, so may be produced singly and need not decay to the LSP.

For HBM2 we increase $s$ up to $10\,$TeV, but many of the features of HBM1 remain unchanged. Again, we have a Higgs boson mass of $124\,$GeV, a light neutralino, with accompanying inert Higgsino to provide the correct Dark Matter abundance, and relatively light squarks, sleptons and gluino. However, for this benchmark we lift the degeneracy in $\kappa_{1,2,3}$ and allow $\kappa_{1,2}$ to be considerably smaller than $\kappa_3$. This scenario therefore cannot be directly matched to one of the points in our scan of Fig.~\ref{tb10s10}. Choosing $\kappa_{1,2}$ small pushes down the mass of the exotic $D$-quarks to $368\,$GeV, allowing them to be pair produced at the LHC via their QCD coupling. For a detailed discussion of this exotic quark production, see Ref.~\cite{Athron:2011wu}, where benchmark C contains exotic quarks with a very similar mass.

One may be concerned that such a light mass for an exotic quark is ruled out by the LHC, however as described in section \ref{exotic-searches} the constraints which have so far been presented are not directly applicable to this case and the detailed studies required to determine limits on our exotic quarks have not yet been carried out. These benchmarks are intended to motivate and aid precisely these urgently needed investigations. Additionally since we obtain the light exotic quarks by setting the $\kappa_{1,2}$ couplings to be very small these can be adjusted to raise the mass of the exotic quarks to $1$ TeV (corresponds to $\kappa_{1,2}\approx 0.055$) without changing the rest of the spectrum by more than $\approx 10\%$.    

We keep these exotic $D$-quarks relatively light also in HBM3, by keeping the same low value of $\kappa_{1,2}$. This scenario has a scalar VEV with $s=20\,$TeV, and one can see that the third generation exotic $D$-quarks, whose mass is fed by a rather large value of $\kappa_3$, become very heavy, over $15\,$TeV.  In HBM4 our scalar VEV becomes very large indeed, with $s=50\,$TeV, but maintains reasonably light first and second generation exotic $D$-quarks, now with $1025\,$GeV, which should still be within reach of the LHC. This is despite now having a large value for $m_0$, and consequently squarks that are way beyond the reach of the LHC.

Finally, we give an example of a benchmark, HBM5, where most of the states are extremely heavy, with $s=100\,$TeV. Here, we only make two concessions towards a light spectrum: firstly, we keep $M_{1/2}$ small, which keeps our two lightest neutralinos and our lightest chargino light, and our gluino relatively light; and secondly, we maintain a small value of $\lambda_1$ to provide a light inert Higgsinos that can satisfy Eq.~(\ref{DM-cond}).
In this scenario we have returned to degenerate $\kappa_i$, so this example is one of the green points in Fig.~\ref{tb10s100}. Without the small value of $\kappa_{1,2}$, the exotic $D$-quarks become extremely heavy (of order $50\,$TeV) well beyond the search reach of the LHC. However, even for supersymmetric scenarios with $s=100\,$TeV we may still have new supersymmetric particles to be discovered at the LHC, since the gaugino sector and a few inert Higgs states are still accessible.

Further studies on these benchmarks will be facilitated by the implementation of the E$_6$SSM into codes like SARAH \cite{Staub:2009bi} and CalcHEP \cite{Pukhov:2004ca}, which are in preparation \cite{Dresden} and also an extension of tools developed for Ref.~\cite{Belyaev:2012si} to include all exotic states in the E$_6$SSM.

\section{Conclusions \label{conclusions}}
In this paper we have studied the parameter space of the constrained exceptional supersymmetric standard model (cE$_6$SSM) consistent with a Higgs signal near 125 GeV and the LHC searches for squarks, gluinos and
$Z^\prime$. The cE$_6$SSM parameter space consistent with correct electroweak symmetry breaking,
is represented by scans in the $(m_0, M_{1/2})$ plane for fixed $Z'$ mass
and $\tan \beta$, with squark, gluino and Higgs masses plotted as contours in this plane.
Although the heaviest Higgs masses are achievable for the largest values of $\lambda$,
EWSB is achieved in the cE$_6$SSM for smaller values of $\lambda$. This is because
EWSB requires reasonably large values of $\kappa_i$ which drive $m_S^2$ negative,
and such values of $\kappa_i$ restrict the values of $\lambda$ that can be achieved consistent with
these couplings remaining perturbative. This means that in practice, the tree-level contribution to the
Higgs mass in the cE$_6$SSM is only slightly larger than in the MSSM, so that a 125 GeV Higgs
mass requires a very large loop contribution, similar to the case of the MSSM. For this reason
we have focussed on values of $\tan \beta = 10$, avoiding the very large values of $\tan \beta $
that may raise other phenomenological issues arising from processes such as $B_s \rightarrow \mu \mu$.

We find that a 125 GeV Higgs mass only arises for a sufficiently large $Z'$ mass, mainly above current limits.
To be precise, the value of $s=5$ TeV
corresponding to $M_{Z'}\sim 2$ TeV only has a very small region of parameter space
consistent with a 125 GeV Higgs boson, although there is a larger region available for a 124 GeV or lighter Higgs
bosons. As expected, heavier Higgs bosons are more easily achieved over large regions of parameter
space for larger values of $s=10-100$ TeV. For each of these cases there are two distinct regions
of the $(m_0, M_{1/2})$ plane consistent with a 125 GeV Higgs boson, where both regions correspond to multi-TeV squark masses, but with one of the regions always extending down to
relatively light gluinos, winos and binos, where the gluinos are typically within reach of
the LHC in the 8 TeV or forthcoming 14 TeV runs. Successful dark matter relic abundance may be achieved
over all the parameter space, assuming a bino-like LSP with a nearby heavier inert Higgsino doublet, about
10 GeV heavier, and decoupled inert singlinos.
This scenario will therefore result in conventional gluino decay signatures
similar to those of the MSSM in the region of parameter space with lighter gluinos and
very heavy squarks and sleptons.  This is similar to the focus point of the MSSM, but with the relic
abundance here resulting from the nearby inert Higgsinos (about 10 GeV heavier than the bino)
which provide the distinguishing phenomenological
prediction of the cE$_6$SSM in this scenario.

A set of benchmark points with a Higgs near 125 GeV has been provided which exemplifies the above features and in addition highlights other features
of phenomenological interest such as exotic $D$ fermions within reach of the LHC.
All these benchmarks also exhibit gluinos, winos and binos and inert Higgsinos, within reach of the
forthcoming runs of the LHC, providing the exciting possibility of SUSY discovery even for
squarks and sleptons outside the range of the LHC. These results show that there is still a vast parameter
space of the cE$_6$SSM to be explored, with heavier squarks and sleptons and lighter gauginos
remaining a firm prediction of the model.

\section*{Acknowledgements}
SFK acknowledges partial support
from the STFC Consolidated ST/J000396/1 and EU ITN grants UNILHC 237920 and INVISIBLES 289442 .
SM is partially supported through the NExT Institute. DJM acknowledges
partial support from the STFC Consolidated Grant ST/G00059X/1. The work of R.N. was
supported by the U.S. Department of Energy under Contract
DE-FG02-04ER41291.  The work of PA is supported by the ARC Centre of Excellence for Particle Physics at the Terascale.
PA would like to thank M. Sch\"onherr, D. St\"ockinger and Tony Williams
for helpful comments and discussions regarding this work.


\begin{thebibliography}{99}
%\cite{ATLAS:2012ae}
\bibitem{ATLAS:2012ae}
  G.~Aad {\it et al.}  [ATLAS Collaboration],
  %``Combined search for the Standard Model Higgs boson using up to 4.9
%fb-1 of pp collision data at sqrt(s) = 7 TeV with the ATLAS detector at the LHC,''
  Phys.\ Lett.\ B {\bf 710} (2012) 49
  [arXiv:1202.1408 [hep-ex]].
  %%CITATION = ARXIV:1202.1408;%%

  %\cite{Chatrchyan:2012tx}
\bibitem{Chatrchyan:2012tx}
S.~Chatrchyan {\it et al.}  [CMS Collaboration],
%``Combined results of searches for the standard model Higgs boson in pp
%collisions at sqrt(s) = 7 TeV,''
Phys.\ Lett.\ B {\bf 710} (2012) 26
arXiv:1202.1488 [hep-ex].


 %
\bibitem{Higgs}
G.~Kane, P.~Kumar, R.~Lu and B.~Zheng,
  %``Higgs Mass Prediction for Realistic String/M Theory Vacua,''
  arXiv:1112.1059 [hep-ph];
I.~Gogoladze, Q.~Shafi and C.~S.~Un,
  %``Higgs Boson Mass from t-b-tau Yukawa Unification,''
  arXiv:1112.2206 [hep-ph];
%\cite{Arbey:2011aa}
%\bibitem{Arbey:2011aa}
  A.~Arbey, M.~Battaglia and F.~Mahmoudi,
  %``Constraints on the MSSM from the Higgs Sector - A pMSSM Study of Higgs Searches, Bs -> mu+ mu- and Dark Matter Direct Detection,''
  arXiv:1112.3032 [hep-ph];
  %%CITATION = ARXIV:1112.3032;%%
%\cite{Arbey:2011ab}
%\bibitem{Arbey:2011ab}
  A.~Arbey, M.~Battaglia, A.~Djouadi, F.~Mahmoudi and J.~Quevillon,
  %``Implications of a 125 GeV Higgs for supersymmetric models,''
  arXiv:1112.3028 [hep-ph];
  %%CITATION = ARXIV:1112.3028;%%
%\cite{Heinemeyer:2011aa}
%\bibitem{Heinemeyer:2011aa}
  S.~Heinemeyer, O.~Stal and G.~Weiglein,
  %``Interpreting the LHC Higgs Search Results in the MSSM,''
  arXiv:1112.3026 [hep-ph];
  %%CITATION = ARXIV:1112.3026;%%
%\cite{Li:2011ab}
%\bibitem{Li:2011ab}
  T.~Li, J.~A.~Maxin, D.~V.~Nanopoulos and J.~W.~Walker,
  %``A Higgs Mass Shift to 125 GeV and A Multi-Jet Supersymmetry Signal: Miracle of the Flippons at the \sqrt{s} = 7 TeV LHC,''
  arXiv:1112.3024 [hep-ph];
  %%CITATION = ARXIV:1112.3024;%%
%\cite{EliasMiro:2011aa}
%\bibitem{EliasMiro:2011aa}
  J.~Elias-Miro, J.~R.~Espinosa, G.~F.~Giudice, G.~Isidori, A.~Riotto and A.~Strumia,
  %``Higgs mass implications on the stability of the electroweak vacuum,''
  arXiv:1112.3022 [hep-ph];
  %%CITATION = ARXIV:1112.3022;%%
%\cite{Baer:2011ab}
%\bibitem{Baer:2011ab}
  H.~Baer, V.~Barger and A.~Mustafayev,
  %``Implications of a 125 GeV Higgs scalar for LHC SUSY and neutralino dark matter searches,''
  arXiv:1112.3017 [hep-ph];
  %%CITATION = ARXIV:1112.3017;%%
%\cite{Englert:2011aa}
%\bibitem{Englert:2011aa}
  C.~Englert, T.~Plehn, M.~Rauch, D.~Zerwas and P.~M.~Zerwas,
  %``LHC: Standard Higgs and Hidden Higgs,''
  arXiv:1112.3007 [hep-ph];
  %%CITATION = ARXIV:1112.3007;%%
%\cite{Kahana:2011aa}
%\bibitem{Kahana:2011aa}
  D.~E.~Kahana and S.~H.~Kahana,
  %``Higgs and Top Masses from Dynamical Symmetry Breaking - Revisited,''
  arXiv:1112.2794 [hep-ph];
  %%CITATION = ARXIV:1112.2794;%%
%\cite{Xing:2011aa}
%\bibitem{Xing:2011aa}
  Z.~-z.~Xing, H.~Zhang and S.~Zhou,
  %``Impacts of the Higgs mass on vacuum stability, running fermion masses and two-body Higgs decays,''
  arXiv:1112.3112 [hep-ph];
  %%CITATION = ARXIV:1112.3112;%%
%\cite{Moroi:2011aa}
%\bibitem{Moroi:2011aa}
  T.~Moroi, R.~Sato and T.~T.~Yanagida,
  %``Extra Matters Decree the Relatively Heavy Higgs of Mass about 125 GeV in the Supersymmetric Model,''
  arXiv:1112.3142 [hep-ph];
  %%CITATION = ARXIV:1112.3142;%%
%\cite{Guo:2011ab}
%\bibitem{Guo:2011ab}
  G.~Guo, B.~Ren and X.~-G.~He,
  %``LHC Evidence Of A 126 GeV Higgs Boson From $H \to \gamma \gamma$ With Three And Four Generations,''
  arXiv:1112.3188 [hep-ph];
  %%CITATION = ARXIV:1112.3188;%%
%\cite{Cheung:2011aa}
%\bibitem{Cheung:2011aa}
  C.~Cheung and Y.~Nomura,
  %``Higgs Descendants,''
  arXiv:1112.3043 [hep-ph];
  %%CITATION = ARXIV:1112.3043;%%
%\cite{Draper:2011aa}
%\bibitem{Draper:2011aa}
  P.~Draper, P.~Meade, M.~Reece and D.~Shih,
  %``Implications of a 125 GeV Higgs for the MSSM and Low-Scale SUSY Breaking,''
  arXiv:1112.3068 [hep-ph];
  %%CITATION = ARXIV:1112.3068;%%
%\cite{Moroi:2011ab}
%\bibitem{Moroi:2011ab}
  T.~Moroi and K.~Nakayama,
  %``Wino LSP detection in the light of recent Higgs searches at the LHC,''
  arXiv:1112.3123 [hep-ph];
  %%CITATION = ARXIV:1112.3123;%%
%\cite{Ferreira:2011aa}
%\bibitem{Ferreira:2011aa}
  P.~M.~Ferreira, R.~Santos, M.~Sher and J.~P.~Silva,
  %``Implications of the LHC two-photon signal for two-Higgs-doublet models,''
  arXiv:1112.3277 [hep-ph];
  %%CITATION = ARXIV:1112.3277;%%
%\cite{Djouadi:2011aa}
%\bibitem{Djouadi:2011aa}
  A.~Djouadi, O.~Lebedev, Y.~Mambrini and J.~Quevillon,
  %``Implications of LHC searches for Higgs--portal dark matter,''
  arXiv:1112.3299 [hep-ph];
  %%CITATION = ARXIV:1112.3299;%%
%\cite{Carena:2011aa}
%\bibitem{Carena:2011aa}
  M.~Carena, S.~Gori, N.~R.~Shah and C.~E.~M.~Wagner,
  %``A 125 GeV SM-like Higgs in the MSSM and the $\gamma \gamma$ rate,''
  arXiv:1112.3336 [hep-ph];
  %%CITATION = ARXIV:1112.3336;%%
  %\cite{Kadastik:2011aa}
%\bibitem{Kadastik:2011aa}
  M.~Kadastik, K.~Kannike, A.~Racioppi and M.~Raidal,
  %``Implications of 125 GeV Higgs boson on scalar dark matter and on the CMSSM phenomenology,''
  arXiv:1112.3647 [hep-ph];
  %%CITATION = ARXIV:1112.3647;%%
%\cite{Akula:2011aa}
%\bibitem{Akula:2011aa}
  S.~Akula, B.~Altunkaynak, D.~Feldman, P.~Nath and G.~Peim,
  %``Higgs Boson Mass Predictions in SUGRA Unification and Recent LHC-7 Results,''
  arXiv:1112.3645 [hep-ph];
  %%CITATION = ARXIV:1112.3645;%%
%\cite{Buchmueller:2011ab}
%\bibitem{Buchmueller:2011ab}
  O.~Buchmueller, R.~Cavanaugh, A.~De Roeck, M.~J.~Dolan, J.~R.~Ellis, H.~Flacher, S.~Heinemeyer and G.~Isidori {\it et al.},
  %``Higgs and Supersymmetry,''
  arXiv:1112.3564 [hep-ph];
  %%CITATION = ARXIV:1112.3564;%%
%\cite{Harlander:2011aa}
%\bibitem{Harlander:2011aa}
%  R.~Harlander, M.~Kramer and M.~Schumacher,
  %``Bottom-quark associated Higgs-boson production: reconciling the four- and five-flavour scheme approach,''
%  arXiv:1112.3478 [hep-ph];
  %%CITATION = ARXIV:1112.3478;%%
%\cite{Cao:2011sn}
%\bibitem{Cao:2011sn}
  J.~Cao, Z.~Heng, D.~Li and J.~M.~Yang,
  %``Current experimental constraints on the lightest Higgs boson mass in the constrained MSSM,''
  arXiv:1112.4391 [hep-ph];
  %%CITATION = ARXIV:1112.4391;%%
%\cite{Strege:2011pk}
%\bibitem{Strege:2011pk}
  C.~Strege, G.~Bertone, D.~G.~Cerdeno, M.~Fornasa, R.~R.~de Austri and R.~Trotta,
  %``Updated global fits of the cMSSM including the latest LHC SUSY and Higgs searches and XENON100 data,''
  arXiv:1112.4192 [hep-ph];
  %%CITATION = ARXIV:1112.4192;%%
%\cite{Burdman:2011ki}
%\bibitem{Burdman:2011ki}
  G.~Burdman, C.~Haluch and R.~Matheus,
  %``Is the LHC Observing the Pseudo-scalar State of a Two-Higgs Doublet Model ?,''
  arXiv:1112.3961 [hep-ph];
  %%CITATION = ARXIV:1112.3961;%%
A.~Arhrib, R.~Benbrik, M.~Chabab, G.~Moultaka and L.~Rahili,
  %``Higgs boson decay into 2 photons in the type~II Seesaw Model,''
  arXiv:1112.5453 [hep-ph];
A.~Bottino, N.~Fornengo and S.~Scopel,
  %``Phenomenology of light neutralinos in view of recent results at the CERN Large Hadron Collider,''
  arXiv:1112.5666 [hep-ph];
%
%\cite{Hall:2011aa}
%\bibitem{Hall:2011aa}
  L.~J.~Hall, D.~Pinner and J.~T.~Ruderman,
  %``A Natural SUSY Higgs Near 126 GeV,''
  arXiv:1112.2703 [hep-ph];
  %%CITATION = ARXIV:1112.2703;%%
%
   %\cite{Arvanitaki:2011ck}
%\bibitem{Arvanitaki:2011ck}
  A.~Arvanitaki and G.~Villadoro,
  %``A Non Standard Model Higgs at the LHC as a Sign of Naturalness,''
  arXiv:1112.4835 [hep-ph];
  %%CITATION = ARXIV:1112.4835;%%
 %\cite{Ellwanger:2011aa}
%\bibitem{Ellwanger:2011aa}
  U.~Ellwanger,
  %``A Higgs boson near 125 GeV with enhanced di-photon signal in the NMSSM,''
  arXiv:1112.3548 [hep-ph];
  %%CITATION = ARXIV:1112.3548;%%
 %\cite{Gunion:2012zd}
%\bibitem{Gunion:2012zd}
  J.~F.~Gunion, Y.~Jiang and S.~Kraml,
  %``The constrained NMSSM and Higgs near 125 GeV,''
  arXiv:1201.0982 [hep-ph];
  %%CITATION = ARXIV:1201.0982;%%
%\cite{King:2012is}
%\bibitem{King:2012is}
  S.~F.~King, M.~Muhlleitner and R.~Nevzorov,
  %``NMSSM Higgs Benchmarks Near 125 GeV,''
  Nucl.\ Phys.\ B {\bf 860} (2012) 207
  [arXiv:1201.2671 [hep-ph]].
  %%CITATION = ARXIV:1201.2671;%%
%\cite{Cao:2012fz}
%\bibitem{Cao:2012fz} 
  J.~Cao, Z.~Heng, J.~M.~Yang, Y.~Zhang and J.~Zhu,
  %``A SM-like Higgs near 125 GeV in low energy SUSY: a comparative study for MSSM and NMSSM,''
  JHEP {\bf 1203}, 086 (2012)
  [arXiv:1202.5821 [hep-ph]].
  %%CITATION = ARXIV:1202.5821;%%.
%\cite{Wang:2012gm}
%\bibitem{Wang:2012gm} 
  L.~Wang and X.~-F.~Han,
  %``The recent Higgs boson data and Higgs triplet model with vector-like quark,''
  arXiv:1206.1673 [hep-ph].
  %%CITATION = ARXIV:1206.1673;%%

%\cite{Fowlie:2012im}
%%\bibitem{Fowlie:2012im} 
  A.~Fowlie, M.~Kazana, K.~Kowalska, S.~Munir, L.~Roszkowski, E.~M.~Sessolo, S.~Trojanowski and Y.~-L.~S.~Tsai,
  %``The CMSSM Favoring New Territories: The Impact of New LHC Limits and a 125 GeV Higgs,''
  arXiv:1206.0264 [hep-ph].
  %%CITATION = ARXIV:1206.0264;%%


%\cite{Djouadi:2005gj}
\bibitem{Djouadi:2005gj}
  A.~Djouadi,
  %``The Anatomy of electro-weak symmetry breaking. II. The Higgs bosons in the minimal supersymmetric model,''
  Phys.\ Rept.\  {\bf 459}, 1-241 (2008).
  [hep-ph/0503173].




    \bibitem{genNMSSM1} P. Fayet, Nucl. Phys. B \textbf{90} (1975) 104; Phys. Lett.
B \textbf{64} (1976) 159; Phys. Lett. B \textbf{69} (1977) 489 and Phys. Lett. B
\textbf{84} (1979) 416; H.P. Nilles, M. Srednicki and D. Wyler, Phys. Lett. B
\textbf{120} (1983) 346; J.M. Frere, D.R. Jones and S. Raby, Nucl. Phys. B
\textbf{222} (1983) 11; J.P. Derendinger and C.A. Savoy, Nucl. Phys. B
\textbf{237} (1984) 307;  A.I. Veselov, M.I. Vysotsky and K.A. Ter-Martirosian,
Sov. Phys. JETP \textbf{63} (1986) 489; J.R. Ellis, J.F. Gunion, H.E. Haber, L.
Roszkowski and F. Zwirner, Phys. Rev. D \textbf{39}  (1989) 844; M. Drees, Int.
J. Mod. Phys. A \textbf{4}  (1989) 3635.
%
\bibitem{genNMSSM2} U. Ellwanger, M. Rausch de Traubenberg and C.A. Savoy, Phys.
Lett. B \textbf{315} (1993) 331, Z. Phys. C {\bf 67} (1995) 665 and Nucl. Phys.
B \textbf{492} (1997) 307; U.~Ellwanger, Phys.\ Lett.\  B {\bf 303} (1993) 271; P.
Pandita, Z. Phys. C \textbf{59} (1993) 575; T. Elliott, S.F. King and P.L.
White, Phys. Rev. D {\bf 49} (1994) 2435; S.F. King and P.L. White, Phys. Rev. D
\textbf{52} (1995) 4183;  F.~Franke and H.~Fraas, Int.\ J.\ Mod.\ Phys.\  A {\bf
12} (1997) 479.

\bibitem{Nevzorov:2004ge}
D.~J.~Miller, R.~Nevzorov and P.~M.~Zerwas,
%``The Higgs sector of the next-to-minimal supersymmetric standard model,''
Nucl.\ Phys.\ B {\bf 681} (2004) 3 [hep-ph/0304049].
%R.~Nevzorov, D.~J.~Miller, {\it Proceedings to the 7th Workshop "What comes beyond
%the Standard Model"}, ed.  by N. S. Mankoc-Borstnik, H. B. Nielsen, C. D. Froggatt,
%D. Lukman, DMFA--Zaloznistvo, Ljubljana, 2004, p. 107 [hep-ph/0411275].

%
%\cite{Ellwanger:2009dp}
\bibitem{Ellwanger:2009dp}
 U.~Ellwanger, C.~Hugonie, A.~M.~Teixeira,
 %``The Next-to-Minimal Supersymmetric Standard Model,''
 Phys.\ Rept.\  {\bf 496} (2010) 1
 [arXiv:0910.1785 [hep-ph]];
U.~Ellwanger,
  %``Higgs Bosons in the Next-to-Minimal Supersymmetric Standard Model at the LHC,''
  Eur.\ Phys.\ J.\ C {\bf 71} (2011) 1782
  [arXiv:1108.0157 [hep-ph]].
  %%CITATION = ARXIV:1108.0157;%%
%

%\cite{King:2012is}
%\bibitem{King:2012is}
\bibitem{King:2012is}
  S.~F.~King, M.~Muhlleitner and R.~Nevzorov,
  %``NMSSM Higgs Benchmarks Near 125 GeV,''
  Nucl.\ Phys.\ B {\bf 860} (2012) 207
  [arXiv:1201.2671 [hep-ph]].
  %%CITATION = ARXIV:1201.2671;%%







%\cite{King:2005jy}
\bibitem{King:2005jy}
  S.~F.~King, S.~Moretti and R.~Nevzorov,
  %``Theory and phenomenology of an exceptional supersymmetric standard model,''
  Phys.\ Rev.\  D {\bf 73}, 035009 (2006)
  [arXiv:hep-ph/0510419].
  %%CITATION = PHRVA,D73,035009;%%
%\cite{King:2005my}
\bibitem{King:2005my}
  S.~F.~King, S.~Moretti and R.~Nevzorov,
  %``Exceptional supersymmetric standard model,''
  Phys.\ Lett.\  B {\bf 634}, 278 (2006)
  [arXiv:hep-ph/0511256].
  %%CITATION = PHLTA,B634,278;%%




%\cite{King:2007uj}
\bibitem{King:2007uj}
  S.~F.~King, S.~Moretti and R.~Nevzorov,
  %``Gauge coupling unification in the exceptional supersymmetric standard model,''
  Phys.\ Lett.\ B {\bf 650} (2007) 57
  [hep-ph/0701064].
  %%CITATION = HEP-PH/0701064;%%




%\cite{Athron:2009bs}
\bibitem{Athron:2009bs}
  P.~Athron, S.~F.~King, D.~J.~Miller, S.~Moretti and R.~Nevzorov,
  %``The Constrained Exceptional Supersymmetric Standard Model,''
  Phys.\ Rev.\  D {\bf 80} (2009) 035009
  [arXiv:0904.2169 [hep-ph]].
  %%CITATION = PHRVA,D80,035009;%%
%\cite{Athron:2009ue}
\bibitem{Athron:2009ue}
  P.~Athron, S.~F.~King, D.~J.~.~Miller, S.~Moretti, R.~Nevzorov and R.~Nevzorov,
  %``Predictions of the Constrained Exceptional Supersymmetric Standard Model,''
  Phys.\ Lett.\  B {\bf 681}, 448 (2009)
  [arXiv:0901.1192 [hep-ph]].
  %%CITATION = PHLTA,B681,448;%%

%\cite{Athron:2011wu}
\bibitem{Athron:2011wu}
  P.~Athron, S.~F.~King, D.~J.~Miller, S.~Moretti and R.~Nevzorov,
  %``LHC Signatures of the Constrained Exceptional Supersymmetric Standard
  %Model,''
  Phys.\ Rev.\  D {\bf 84}, 055006 (2011)
  [arXiv:1102.4363 [hep-ph]].
  %%CITATION = PHRVA,D84,055006;%%

%%%%%%%%%%%%%%%%%%%%%%%cE6SSM-section%%%%%%%%%%%%%%%%%%%%%%%%%%%%%%%%%%%%%%%%%%

\bibitem{Accomando:2006ga}
S.~F.~King, S.~Moretti, R.~Nevzorov,
%``Spectrum of Higgs particles in the ESSM,''
arXiv:hep-ph/0601269;
S. Kraml {\it et al.} (eds.), {\it Workshop on CP studies and
non-standard Higgs physics}, CERN--2006--009, hep-ph/0608079;
S.~F.~King, S.~Moretti, R.~Nevzorov,
%``E(6)SSM,''
AIP Conf.\ Proc.\  {\bf 881} (2007) 138;
%[arXiv:hep-ph/0610002];
R.~Howl, S.~F.~King,
%``Minimal E_6 Supersymmetric Standard Model,''
JHEP {\bf 0801} (2008) 030;
P.~Athron, S.~F.~King, D.~J.~Miller, S.~Moretti, R.~Nevzorov,
%``The Constrained E$_6$SSM,''
arXiv:0810.0617 [hep-ph].

\bibitem{Howl:2008xz}
R.~Howl, S.~F.~King,
%``Exceptional Supersymmetric Standard Models with non-Abelian Discrete Family
%Symmetry,''
JHEP {\bf 0805} (2008) 008.

\bibitem{King:2008qb}
S.~F.~King, R.~Luo, D.~J.~Miller, R.~Nevzorov,
%``Leptogenesis in the Exceptional Supersymmetric Standard Model: flavour
%dependent lepton asymmetries,''
JHEP {\bf 0812} (2008) 042.

%\cite{Hall:2011zq}
\bibitem{Hall:2011zq}
J.~P.~Hall and S.~F.~King,
%``Bino Dark Matter and Big Bang Nucleosynthesis in the Constrained $E_6SSM with Massless Inert Singlinos,''
JHEP {\bf 1106} (2011) 006
[arXiv:1104.2259 [hep-ph]].
%%CITATION = ARXIV:1104.2259;%%

%\cite{Nevzorov:2001um}
\bibitem{Nevzorov:2001um}
P.~A.~Kovalenko, R.~B.~Nevzorov and K.~A.~Ter-Martirosian,
%``Masses of Higgs bosons in supersymmetric theories,''
Phys.\ Atom.\ Nucl.\  {\bf 61} (1998) 812
[Yad.\ Fiz.\  {\bf 61} (1998) 898];
R.~B.~Nevzorov and M.~A.~Trusov,
%``Particle spectrum in the modified NMSSM in the strong Yukawa coupling
%limit. (In Russian),''
J.\ Exp.\ Theor.\ Phys.\  {\bf 91} (2000) 1079
[Zh.\ Eksp.\ Teor.\ Fiz.\  {\bf 91} (2000) 1251]
[arXiv:hep-ph/0106351];
R.~B.~Nevzorov, K.~A.~Ter-Martirosyan and M.~A.~Trusov,
%``Higgs bosons in the simplest SUSY models,''
Phys.\ Atom.\ Nucl.\  {\bf 65} (2002) 285
[Yad.\ Fiz.\  {\bf 65} (2002) 311]
[arXiv:hep-ph/0105178].

%\cite{Miller:2005qua}
\bibitem{Miller:2005qua}
C. Panagiotakopoulos, A. Pilaftsis, Phys. Rev. D {\bf 63} (2001) 055003;
D.~J.~Miller, R.~Nevzorov
%``The Peccei-Quinn axion in the next-to-minimal supersymmetric standard model,''
[arXiv:hep-ph/0309143];
R. Nevzorov, D. J.  Miller, {\it Proceedings to the 7th Workshop "What comes beyond
the Standard Model"}, ed.  by N. S. Mankoc-Borstnik, H. B. Nielsen, C. D. Froggatt,
D. Lukman, DMFA--Zaloznistvo, Ljubljana, 2004, p. 107; hep-ph/0411275;
D. J. Miller, S. Moretti, R. Nevzorov, {\it Proceedings to the 18th International
Workshop on High-Energy Physics and Quantum Field Theory (QFTHEP 2004)},
ed. by M.N. Dubinin, V.I. Savrin, Moscow, Moscow State Univ., 2004. p. 212; hep-ph/0501139.

\bibitem{42}
J.~Rich, M.~Spiro, J.~Lloyd--Owen, Phys.\ Rept.\  {\bf 151} (1987) 239;
P.~F.~Smith, Contemp.\ Phys.\ {\bf 29} (1988) 159;
T.~K.~Hemmick et al., Phys.\ Rev.\  D {\bf 41} (1990) 2074.

%\cite{Allanach:2001kg}
\bibitem{Allanach:2001kg}
B.~C.~Allanach,
%``SOFTSUSY: A C++ program for calculating supersymmetric spectra,''
Comput.\ Phys.\ Commun.\  {\bf 143}  (2002) 305.
%[arXiv:hep-ph/0104145].
%%CITATION = CPHCB,143,305;%%


%%%%%%%%%%%%%%%%%%%%%%%%%%%%%%%%%%%%%%%%%%%%%%%%%%%%%%%%%%%%%%%%%%%%%%%%%%%%%%%

%\cite{Moretti:1994ds}
\bibitem{Moretti:1994ds}
  S.~Moretti and W.~J.~Stirling,
  %``Contributions of below threshold decays to MSSM Higgs branching ratios,''
  Phys.\ Lett.\  B {\bf 347} (1995) 291
  [Erratum, {\it ibidem}  {\bf 366} (1996) 451].
%  [arXiv:hep-ph/9412209].
  %%CITATION = PHLTA,B347,291;%%


%\cite{Kunszt:1996yp}
\bibitem{Kunszt:1996yp}
  Z.~Kunszt, S.~Moretti and W.~J.~Stirling,
  %``Higgs production at the LHC: An Update on cross-sections and branching
  %ratios,''
  Z.\ Phys.\  C {\bf 74} (1997) 479
  [arXiv:hep-ph/9611397].
  %%CITATION = ZEPYA,C74,479;%%

\bibitem{ATLAS_susy1} The ATLAS Collaboration, ATLAS-CONF-2012-033.
\bibitem{ATLAS_susy2} The ATLAS Collaboration, ATLAS-CONF-2012-037.
\bibitem{ATLAS_susy3} The ATLAS Collaboration, ATLAS-CONF-2012-041.

\bibitem{CMS_razor} The CMS Collaboration, % “Search for supersymmetry with the razor variables at CMS”,
CMS PAS SUS-12-005 (2012).

\bibitem{CMS_MT2} The CMS Collaboration, % “Search for Supersymmetry in hadronic final states using MT2 with the CMS detector at 7 TeV”,
CMS PAS SUS-12-002 (2012).

\bibitem{LeCompte:2011cn}
  T.~J.~LeCompte and S.~P.~Martin,
  %``Large Hadron Collider reach for supersymmetric models with compressed mass spectra,''
  Phys.\ Rev.\ D {\bf 84}, 015004 (2011)
  [arXiv:1105.4304 [hep-ph]].
  %%CITATION = ARXIV:1105.4304;%%

\bibitem{Aad:2011cw}
  G.~Aad {\it et al.}  [ATLAS Collaboration],
  %``Search for scalar bottom pair production with the ATLAS detector in pp Collisions at sqrt{s} = 7 TeV,''
  Phys.\ Rev.\ Lett.\  {\bf 108} (2012) 181802
  [arXiv:1112.3832 [hep-ex]].
  %%CITATION = ARXIV:1112.3832;%%

\bibitem{ATLAS_stop} The ATLAS Collaboration, %``Search for Scalar Top Quark Pair Production in Natural Gauge Mediated Supersymmetry Models with the ATLAS Detector in pp Collisions at s√=~7~TeV''
ATLAS-CONF-2012-036

\bibitem{Alves:2011wf}
  D.~Alves {\it et al.}  [LHC New Physics Working Group Collaboration],
  %``Simplified Models for LHC New Physics Searches,''
  arXiv:1105.2838 [hep-ph].
  %%CITATION = ARXIV:1105.2838;%%

\bibitem{CMS_simp} The CMS Collaboration, %Interpretation of Searches for Supersymmetry with
Simplified Models",
CMS PAS SUS-11-016 (2011);
% “Search for new physics in the multijets + missing transverse energy final state”,
CMS PAS SUS-12-011 (2012).

\bibitem{ATLAS_chargino} The ATLAS Collaboration, %``Search for supersymmetry in events with three leptons and missing transverse momentum in √s = 7 TeV pp collisions with the ATLAS detector'',
ATLAS-CONF-2012-023



%\cite{Aad:2011uv}
\bibitem{Aad:2011uv}
G.~Aad {\it et al.}  [ATLAS Collaboration],
%``Search for pair production of first or second generation leptoquarks in proton-proton collisions
%at sqrt(s)=7 TeV using the ATLAS detector at the LHC,''
Phys.\ Rev.\ D {\bf 83} (2011) 112006
[arXiv:1104.4481 [hep-ex]].

%\cite{90}
\bibitem{90}
S.~L.~Cheung,
%``Search for Dijet Resonances in $\sqrt{s}$ = 7 TeV Proton-Proton Collisions with the ATLAS Detector at the LHC,''
CERN-THESIS-2011-162.

%\cite{Nevzorov:2012hs}
\bibitem{Nevzorov:2012hs}
R.~Nevzorov,
%``E6 inspired SUSY models with exact custodial symmetry,''
arXiv:1205.5967 [hep-ph].



\bibitem{ATLAS-Zp} The ATLAS Collaboration, ATLAS-CONF-2012-007.
\bibitem{CMS-Zp} The CMS Collaboration, CMS PAS EXO-11-019.
\bibitem{:2012it} 
  S.~Chatrchyan {\it et al.}  [CMS Collaboration],
  %``Search for narrow resonances in dilepton mass spectra in pp collisions at sqrt(s) = 7 TeV,''
  arXiv:1206.1849 [hep-ex].
  %%CITATION = ARXIV:1206.1849;%%
\bibitem{Accomando:2010fz}
  E.~Accomando, A.~Belyaev, L.~Fedeli, S.~F.~King and C.~Shepherd-Themistocleous, %``Z' physics with early LHC data,''
  Phys.\ Rev.\ D {\bf 83}, 075012 (2011)
  [arXiv:1010.6058 [hep-ph]].
  %%CITATION = ARXIV:1010.6058;%%

  %\cite{Kalinowski:2008iq}
\bibitem{Kalinowski:2008iq}
  J.~Kalinowski, S.~F.~King, J.~P.~Roberts,
  %``Neutralino Dark Matter in the USSM,''
  arXiv:0811.2204 [hep-ph].
  %%CITATION = ARXIV:0811.2204;%%

%\cite{Hall:2009aj}
\bibitem{Hall:2009aj}
  J.~P.~Hall and S.~F.~King,
  %``Neutralino Dark Matter with Inert Higgsinos and Singlinos,''
  arXiv:0905.2696 [hep-ph].
  %%CITATION = ARXIV:0905.2696;%%

%\cite{Hall:2010ix}
\bibitem{Hall:2010ix}
J.~P.~Hall, S.~F.~King, R.~Nevzorov, S.~Pakvasa, M.~Sher,
%``Novel Higgs Decays and Dark Matter in the E(6)SSM,''
Phys.\ Rev.\ D {\bf 83} (2011) 075013
[arXiv:1012.5114 [hep-ph]].
%%CITATION = ARXIV:1012.5114;%%

  %\cite{Belyaev:2012si}
\bibitem{Belyaev:2012si}
  A.~Belyaev, J.~P.~Hall, S.~F.~King and P.~Svantesson,
  %``Novel gluino cascade decays in $E_6$ inspired models,''
  arXiv:1203.2495 [hep-ph].
  %%CITATION = ARXIV:1203.2495;%%

  %\cite{King:2012wg}
\bibitem{King:2012wg}
  S.~F.~King and A.~Merle,
  %``Warm Dark Matter from keVins,''
  arXiv:1205.0551 [hep-ph].
  %%CITATION = ARXIV:1205.0551;%%


\bibitem{Staub:2009bi} 
  F.~Staub,
  %``From Superpotential to Model Files for FeynArts and CalcHep/CompHep,''
  Comput.\ Phys.\ Commun.\  {\bf 181}, 1077 (2010)
  [arXiv:0909.2863 [hep-ph]].
  %%CITATION = ARXIV:0909.2863;%%
\bibitem{Pukhov:2004ca} 
  A.~Pukhov,
  %``CalcHEP 2.3: MSSM, structure functions, event generation, batchs, and generation of matrix elements for other packages,''
  hep-ph/0412191.
  %%CITATION = HEP-PH/0412191;%%
\bibitem{Dresden} 
P.~Diessner, G.~Hellwig, G.M.~Pruna, D.~Stockinegr, A.~Voigt, in preparation.


%%\cite{Hall:2011zq}
%\bibitem{Hall:2011zq}
%J.~P.~Hall and S.~F.~King,
%%``Bino Dark Matter and Big Bang Nucleosynthesis in the Constrained $E_6SSM with Massless Inert Singlinos,''
%JHEP {\bf 1106} (2011) 006
%[arXiv:1104.2259 [hep-ph]].
%%%CITATION = ARXIV:1104.2259;%%



\end{thebibliography}
\end{document}